\documentclass[a4,10pt, onecolumn, usenatbib]{mn2e}
\usepackage[dvips]{graphicx, psfrag}
\usepackage{times}
\usepackage{color}
\usepackage{amsmath,amsfonts,amssymb}
\usepackage{textcomp}

\usepackage{natbib}
\usepackage{aas_macros}

\title[Magnetic fields throughout magnetar]
{Magnetic field configurations of {\bf a} magnetar throughout its 
interior and exterior -- core, crust and magnetosphere}

\author[K. Fujisawa \& S. Kisaka]{Kotaro Fujisawa$^{1}$ 
\thanks{E-mail: fujisawa@heap.phys.waseda.ac.jp} 
\thanks{Formerly at 
Department of Earth Science and Astronomy,
Graduate School of Arts and Sciences, University of Tokyo,
Komaba, Meguro-ku, Tokyo 153-8902, Japan 
} 
\&
Shota Kisaka$^{2}$ 
\thanks{Formerly at 
Institute for Cosmic Ray Research, University of Tokyo, 5-1-5 Kashiwa-no-ha, Kashiwa city, Chiba 277-8582, Japan
} 
\\
$^1$Advanced Research Institute for Science and Engineering, 
Waseda University, 3-4-1 Okubo, Shinjuku-ku, Tokyo 169-8555, Japan
\\
$^2$ Institute of Particle and Nuclear Studies, KEK, 
1-1 Oho, Tsukuba 305-0801, Japan
}

\date{
Accepted 2014 September 10. Received 2014 August 25; in original form 2013 November 30}

\def\Vec#1{\mbox{\boldmath $#1$}}
\def\D#1#2{\dfrac{d #1}{d #2}}

\def\P#1#2{\dfrac{\partial #1}{\partial #2}}
\def\PP#1#2{\dfrac{\partial^2 #1}{\partial #2^2}}

\usepackage{multicol}	
\begin{document}

\maketitle

\begin{abstract}
We obtained the magnetic field configurations, 
including both poloidal and toroidal components,
throughout the interior and exterior of magnetars
using a realistic equation of state.
We divided the magnetized star into the 
hydromagnetic equilibrium core, 
Hall equilibrium crust and twisted force-free magnetosphere.
We systematically and simultaneously calculated
these regions under various boundary 
conditions using the Green function relaxation method,
and noted the following interesting characteristics of these 
numerical results. First, the strength and 
structure of core magnetic fields affect 
the crustal magnetic fields.
Second, the current sheet on the core-crust interface 
affects both internal and external magnetic field configurations.
Third, the twisted magnetosphere makes 
a cross-point of magnetic field lines, such as X-point geometry, in the magnetosphere.
The X-point geometry appears and disappears 
according to  the strength of the twisted field in the magnetosphere or
the core-crust boundary conditions. 
Our results mean that both 
Hall magnetohydrodynamics secular evolution and magnetospheric dynamical evolution
are deeply affected by conditions of 
another region and the core-crust stress of magnetars.
\end{abstract}

\begin{multicols}{1} 

\section{Introduction}

Neutron stars have the strongest magnetic fields 
among the stars in the universe. 
In particular, Anomalous X-ray Pulsars (AXPs) 
and Soft Gamma-ray Repeaters (SGRs) are 
considered to be  magnetars, a special class of magnetized neutron stars.
Magnetars have very strong dipole magnetic fields 
whose typical values reach approximately $10^{15}$ G at their surfaces.
They display stationary intense emissions and
dynamical flares  releasing magnetic energy (\citealt{Thompson_Duncan_1995}).
Such magnetic activities are also considered to be their heating source (\citealt{Pons_et_al_2007}).
The magnetic field configurations are important for such 
astrophysical events because the 
decay of magnetic energy deeply depends on their structure.
Especially in relation to flares, both exterior and interior magnetic 
field structures are considered to be essential by many studies 
 (e.g. \citealt{Thompson_Duncan_2001}; 
\citealt{Thompson_Lyutikov_Kulkarni_2002}; 
\citealt{Beloborodov_Thompson_2007};
\citealt{Masada_Nagataki_Shibata_Terasawa_2010}).
Recent studies have performed numerical simulations of crustal Hall magnetohydrodynamics (MHD) 
secular evolutions (\citealt{Perna_Pons_2011}) and strongly twisted force-free
magnetospheric dynamical evolutions (\citealt{Parfrey_Beloborodov_Hui_2013}) 
to understand the flares.
However, since the magnetic field structures of the core-crust 
(interior) and magnetosphere (exterior) should be coupled, 
we must consider them simultaneously.
As a first step, in this paper we will systematically and simultaneously 
calculate the equilibrium magnetic 
field structure of magnetars throughout their interior and exterior.

The Hall drift within the crust is considered to be vital
for the internal field structure of magnetars.
The Hall drift itself does not dissipate 
the magnetic fields, but produces higher order 
components of the magnetic field during the Hall cascade process.  
Such higher order components of the magnetic field promote 
Ohmic dissipation more efficiently  
(\citealt{Jones_1988}; 
\citealt{Goldreich_Reisenegger_1992}; \citealt{Naito_Kojima_1994};
\citealt{Urpin_Shalybkov_1995}; \citealt{Shalybkov_Urpin_1997}; \citealt{Geppert_Rheinhardt_2002};
\citealt{Rheinhardt_Geppert_2002}; \citealt{Rheinhardt_Konekov_Geppert_2004};
\citealt{Hollerbach_Rudiger_2002}; \citealt{Hollerbach_Rudiger_2004};
\citealt{Cumming_et_al_2004}; \citealt{Reisenegger_et_al_2007}; \citealt{Pons_Geppert_2007}).
The timescale of the Hall drift is determined by the strength of the magnetic 
fields within the crust.
The Hall drift is characterized by the magnetic Reynolds number 
${\cal R}_{\rm{m}}$ (\citealt{Pons_Geppert_2007}) as follows:,
\begin{eqnarray}
 {\cal R}_{\rm{m}} \equiv \frac{t_{\rm{Ohm}}}{t_{\rm{Hall}}} = \frac{\sigma B}{e c n_e} 
\sim 10^{3} \left(\frac{B}{10^{15} \mathrm{G}}\right) 
\left(\frac{\sigma}{10^{25} \mathrm{s}^{-1}}  \right),
\label{Eq:R_m}
\end{eqnarray}
where $t_{\rm{Ohm}}$ and $t_{\rm{Hall}}$ are timescales of  Hall drift and 
Ohmic dissipation, respectively.
$\sigma$, $c$, $n_e$, and $e$ are, respectively, the electrical conductivity of the crust,
the speed of light, the electron number density,
the charge of an electron. 
This value reaches approximately 1000 locally
within the magnetar crust, because the crustal magnetic fields of the magnetar 
are sufficient large ($\geq 10^{15}$G).
The Hall drift becomes very effective in the magnetar crust
and its timescale is faster than or comparable to the lifetime of magnetars.
Hall MHD numerical simulations have been 
performed (\citealt{Kojima_Kisaka_2012};  
\citealt{Vigano_Pons_2012}; \citealt{Vigano_et_al_2012}; 
\citealt{Vigano_et_al_2013}; \citealt{Gourgouliatos_Cumming_2014}) and these works 
 examined the Hall drift during the secular timescale. 

As shown in recent works, the toroidal magnetic fields decrease rapidly during the Hall drift
timescale because they are changed into higher order poloidal components by the
Hall cascade, and Ohmic dissipation is promoted by the higher order components
(\citealt{Kojima_Kisaka_2012}; \citealt{Vigano_et_al_2013}; \citealt{Gourgouliatos_Cumming_2014}).
{According to these simulations,
the Hall cascade becomes very effective when 
the initial energy of the toroidal magnetic field is much larger than that of the  poloidal magnetic field.
Therefore the energy ratio of the toroidal to the total magnetic field
is a key factor in the secular evolution of the magnetic fields of magnetars.
To investigate the magnetic field configurations of magnetars in equilibrium, 
we should include the Hall drift.
One approach is the Hall equilibrium study. 

\cite{Gourgouliatos_et_al_2013} calculated 
Hall equilibria with both poloidal and toroidal magnetic fields
by solving the Grad-Shafranov equation.  
They considered crustal magnetic fields in a Hall equilibrium 
and assumed a  vacuum exterior.
As \cite{Gourgouliatos_et_al_2013} pointed out, 
the Hall equilibrium state is similar to
the MHD equilibrium state which has been studied 
for sixty years since Chandrasekhar and his colleagues 
 pioneering works (\citealt{Chandrasekhar_Fermi_1953}; \citealt{Ferraro_1954};
 \citealt{Chandrasekhar_1956}; \citealt{Chandrasekhar_Prendergast_1956};
 \citealt{Prendergast_1956};
 \citealt{Woltjer_1959a}; \citealt{Woltjer_1959b};
 \citealt{Wentzel_1961}; \citealt{Ostriker_Hartwick_1968};
 \citealt{Miketinac_1973}; \citealt{Miketinac_1975};
 \citealt{Bocquet_et_al_1995}; \citealt{Konno_Obata_Kojima_1999};
 \citealt{Ioka_Sasaki_2004}; \citealt{Kiuchi_Yoshida_2008}; \citealt{Kiuchi_Kotake_2008}; 
 \citealt{Haskell_et_al_2008}; \citealt{Lander_Jones_2009}; 
 \citealt{Ciolfi_et_al_2009}; \citealt{Ciolfi_et_al_2010}; \citealt{Duez_Mathis_2010};
 \citealt{Fujisawa_Yoshida_Eriguchi_2012}; 
 \citealt{Fujisawa_et_al_2013};  \citealt{Fujisawa_Eriguchi_2013};
 \citealt{Ciolfi_Rezzolla_2013}).
 Both analytical and theoretical methods for the MHD 
 equilibrium state were developed and investigated in these studies. 
 Recently, some works considered more realistic 
 and complex physical conditions in neutron star and magnetar interiors. 
 \cite{Yoshida_Kiuchi_Shibata_2012}
 considered stratification by chemical potentials 
 in the neutron star interior and obtained
 stably stratified magnetized stars in general relativistic equilibrium.
 \cite{Lander_Andersson_Glampedakis_2012}
 and \cite{Glampedakis_Andersson_Lander_2012} calculated
 magnetized neutron star equilibria with stratification and type II
 superconductivity in a Newtonian framework. 
 \cite{Lander_2013a} and \cite{Lander_2014} solved the superconducting 
 flux tube tension and obtained magnetic field configurations of a neutron star
 with a superconducting core and a normal MHD crust by using a self-consistent field method.
 \cite{Glampedakis_Lander_Andersson_2014} have studied 
 MHD magnetized star in equilibrium with twisted force-free magnetosphere 
by using a self-consistent field method. In their model,  the interior field 
 is in equilibrium with its magnetosphere. 
However, they did not employ a realistic equation of state (EOS)
nor distinguish between its core and crust.
Since the previous Hall MHD equilibrium (\citealt{Gourgouliatos_et_al_2013}) 
study did not include a twisted magnetosphere, 
nobody has obtained magnetized equilibria throughout a core-crust-magnetosphere,
 which is essential for the progress of magnetar understanding.

In this paper, we extended the above  works
and systematically and simultaneously obtained the magnetic field structures across
the core, crust, and magnetosphere for the first time. 
We used a realistic EOS to ensure that the core and crust are treated
in an appropriate manner.  We assumed crustal magnetic fields in the Hall equilibrium, 
because we considered a magnetar with a strong magnetic field where
  the Hall drift would be very effective within its crust.
For simplicity, we assumed that the core is in MHD equilibrium.
As described, the formulation of the Hall equilibrium state is very similar 
to the MHD equilibrium state (\citealt{Gourgouliatos_et_al_2013}), 
but their physical meanings are different from each other.
The MHD equilibrium state depends on  mass density profiles
because it describes the matter force balance.
On the other hand, the Hall equilibrium state in the crust 
is determined by the electron number density distribution within the crust only
and neglects the force balance of the crust. 
Therefore, the crust in this paper is elastic and stressed 
by Lorentz force of the crustal magnetic fields.  
In this paper, we  assume that the magnetosphere is twisted and force-free.
We adopted the equatorial shearing and ring models
of \cite{Parfrey_Beloborodov_Hui_2013} as  magnetospheric models.
Since we are interested in the magnetosphere near the star,
we neglected the rotation of the magnetosphere.
In summary, we consider a MHD equilibrium core, Hall equilibrium crust and 
twisted force-free magnetosphere in this paper.
This paper is organized as follows. In section 2, the basic equations
and models are described. In section 3, we show the numerical results.
We will provide discussion and our conclusion in section 4.

\section{Formulation and models}

\subsection{Basic equations and integral forms}
\label{Sec:Basic_equation}

We consider the MHD equilibrium core, Hall equilibrium crust, and twisted force-free 
magnetosphere simultaneously in this paper. 
We use both spherical coordinates ($r$, $\theta$, $\varphi$) and 
cylindrical coordinates ($R$, $\varphi$, $z$).
We assume that the system is stationary and axisymmetric.
The magnetic fields are defined by two scalar potentials, $\Psi$ and $I$, as
\begin{eqnarray}
\Vec{B} = \frac{1}{r \sin \theta} (\nabla \Psi \times \Vec{e}_\varphi) + \frac{I}{r \sin \theta} \Vec{e}_\varphi,
\end{eqnarray}
where $\Psi$ is a poloidal magnetic flux function and $I$ is a poloidal current density flux function.
These magnetic fields components are related to the current density through the Maxwell Amp\`ere equation as follows:
\begin{eqnarray}
4 \pi \frac{\Vec{j}}{c} = \nabla \times \Vec{B} = \frac{1}{r \sin \theta} (\nabla I \times \Vec{e}_\varphi) 
 - \frac{1}{r \sin \theta} \Delta^{*} \Psi \Vec{e}_\varphi,
\end{eqnarray}
where $\Delta^*$ is the Grad-Shafranov operator below:
\begin{eqnarray}
\Delta^* =  \PP{}{r} + \frac{\sin \theta}{r^2} \P{}{\theta }\left(\frac{1}{\sin \theta} \P{}{\theta} \right).
\end{eqnarray}
To obtain the magnetized equilibria, we need to solve the following elliptic type equation:
\begin{eqnarray}
\Delta^* \Psi = - 4 \pi r \sin \theta \frac{j_\varphi}{c}.
\label{Eq:GS_diff}
\end{eqnarray}
The right hand side of this equation contains the toroidal current density as a source term.
This is derived from the matter equation in each region.

The Hall equilibrium state within the crust is described by the 
Hall equilibrium equation (\citealt{Gourgouliatos_et_al_2013} and App. \ref{App:eqs}):
\begin{eqnarray}
 \nabla \times \Big[\frac{c}{4\pi e n_e} \Vec{B} \times (\nabla \times \Vec{B}) \Big] = 0.
\end{eqnarray}
This equation results in the following functional form of toroidal current density: 
\begin{eqnarray}
4 \pi \frac{j_\varphi}{c} = \frac{I(\Psi) I'(\Psi)}{r\sin \theta} + 4 \pi n_e r \sin \theta S(\Psi),
\label{Eq:jphi_Hall}
\end{eqnarray}
where $I(\Psi)$ and $S(\Psi)$ are arbitrary functions of $\Psi$, and
$I' = \D{I}{\Psi}$. The crustal toroidal current density is described by this equation.

As \cite{Gourgouliatos_et_al_2013} pointed out, the toroidal 
current density in the Hall equilibrium system 
is similar to that of a barotropic MHD equilibrium system. 
The stationary MHD Euler equation without rotation and meridional flow is described as
\begin{eqnarray}
 \frac{1}{\rho} \nabla p = - \nabla \phi_g + \frac{1}{\rho} \left(\frac{\Vec{j}}{c} \times \Vec{B} \right),
\end{eqnarray}
where $\rho$, $p$, and $\phi_g$ are the mass density, pressure and gravitational potential, respectively.
From the integrability condition of the Euler equation, we can obtain the relation
\begin{eqnarray}
 4\pi\frac{j_\varphi}{c} = \frac{I(\Psi)I'(\Psi)}{r\sin \theta} + 4 \pi \rho r \sin \theta F(\Psi),
\label{Eq:jphi_MHD}
\end{eqnarray}
where $F(\Psi)$ is an another arbitrary function of $\Psi$. 
This equation expresses the core toroidal current density.

The twisted magnetosphere without rotation satisfies the force-free condition as
\begin{eqnarray}
 \frac{\Vec{j}}{c} \times \Vec{B} = 0.
\end{eqnarray}
Using this condition, we derived the functional form of toroidal current density as follows: 
\begin{eqnarray}
 4\pi \frac{j_\varphi}{c} = \frac{I(\Psi)I'(\Psi)}{r \sin \theta}.
\end{eqnarray}
The toroidal current density in the magnetosphere is only described by the arbitrary function $I(\Psi)$. 

We can calculate the magnetized equilibria using these functional forms of $j_\varphi$
throughout the star. In order to easily include the boundary conditions, 
we calculated the integrated form of Eq. (\ref{Eq:GS_diff}) using the Green function (see App. \ref{App:Legendre}),
\begin{eqnarray}
 \frac{\Psi(\Vec{r})}{r \sin \theta} \sin \varphi =  
 \frac{1}{c}\int_V \frac{j_\varphi(\Vec{r'})}{|\Vec{r}-\Vec{r}'|} \sin \varphi' dV' + h.ts ,
\label{Eq:Green_function}
\end{eqnarray}
where $h.ts$ denotes the homogeneous terms of the Laplacian as shown below:
\begin{eqnarray}
 h.ts = \sum_{n=1}^{\infty} \Big(a_n r^{n} P_n^1 (\cos \theta) + b_n r^{-n-1} P_n^1 (\cos \theta) \Big) \sin \varphi,
\end{eqnarray}
where $P_n^1 (\cos \theta)$ are associated Legendre functions. The coefficients $a_n$ and $b_n$ are determined by 
the boundary conditions of $\Psi$.  These 
homogeneous terms come from current sheets 
on the boundaries (see \citealt{Fujisawa_Eriguchi_2013} and App. \ref{App:analytical}).

The physical dimension of $F(\Psi)$ is different from $S(\Psi)$ because the 
dimensions of $\rho$ and $n_e$ differ (\citealt{Gourgouliatos_et_al_2013}). 
Therefore, the functional form of $F(\Psi)$ can differ from $S(\Psi)$.
On the other hand, the physical dimension of $I(\Psi)$ in each region
 is the same because this conserved quantity is only 
obtained from the axisymmetry condition (see App. \ref{App:eqs}).
We used the same functional form of $I(\Psi)$ in the core and crust regions.

\subsection{Models of the internal magnetic field}

We produced four models of the internal magnetic field
according to the core-crust toroidal current density and boundary conditions.
We must fix the functional forms of $I(\Psi)$, $F(\Psi)$, and $S(\Psi)$, and the 
current sheet on the core-crust boundary in order to achieve these models.
The magnetic field configuration of each model is 
displayed in Fig. \ref{Fig:models}.
Each model is described as follows:
\begin{enumerate}
 \item 
Model I is a purely crustal open magnetic field in a 
Hall equilibrium. Both poloidal and toroidal magnetic fields 
satisfy the Hall equilibrium state.
This model is equivalent to the configurations used by  \cite{Gourgouliatos_et_al_2013}.
This model requires the opposite current sheet on the core-crust boundary
to prevent the poloidal magnetic fields from entering the core (see App. \ref{App:analytical}). 
The inner boundary conditions in this case are $\Psi = 0$ and $I = 0$ on the core-crust interface.

\item
Model II is a purely crustal current model. The poloidal magnetic fields can
penetrate the core region, but in this model, the toroidal magnetic field is confined within the 
crust region. The current density only exists within the crust.
This model does not have the opposite current sheet on the core-crust boundary.
Therefore, the core magnetic field is the inner vacuum solution of the crustal current. 
The inner boundary conditions in this model are $\Psi \neq 0$ and $I = 0$ on the core-crust interface.

\item
Model III is a core-crust current model. Both poloidal and toroidal magnetic 
fields can exist in the core and crust regions. In other words, 
both poloidal and toroidal current densities can exist in the core and crust regions.
In this model, the toroidal current density in the core flows in the same direction as the crust current.
 The inner boundary conditions in this model are $\Psi \neq 0$ and $I \neq 0$ on the core-crust interface.

\item
Model IV is a core-crust current model.
The toroidal current density in the core flows in the opposite direction 
to the crustal current in this model. 
The core magnetic fields' energy in this model is
 much larger than the crustal magnetic fields' energy, 
because such oppositely flowing toroidal current density 
makes the core magnetic fields' energy much larger 
(\citealt{Fujisawa_Eriguchi_2013}). 
The inner boundary conditions in this model are $\Psi \neq 0$ 
and $I \neq 0$ on the core-crust interface.
\end{enumerate}

We calculated analytical solutions of these models without toroidal
magnetic fields (see App. \ref{App:analytical}). 
Models III and IV can have arbitrary current sheets on the core-crust boundaries.
Such current sheets cause the arbitrary magnetic pressure on the bottom of the crust
(\citealt{Braithwaite_Spruit_2006}). 
We only calculated dipole current sheets 
in these models, but are also able to calculate the arbitrary higher 
order current sheet (see details of the current sheet in 
\citealt{Fujisawa_Eriguchi_2013}).
The explicit form of the dipole toroidal current sheet $j_s$ is 
\begin{eqnarray}
 j_s = j_0 \delta (r - r_{\mathrm{in}}) \sin \theta,
\label{Eq:j_sur}
\end{eqnarray}
where $r_{\rm in}$ denotes the radius of crust-core interface
and $j_0$ is the  strength of the current density. $\delta$ is
Dirac's delta function. This current sheet makes dipole magnetic fields 
(see the analytical solutions in App. \ref{App:analytical}).
We calculated   solutions of model IV with and without current sheets 
in order to examine the influence of the current sheet.
We calculated  solutions of model IV with both positive and negative 
current sheets. Therefore, a total of 6 solution types 
for the models of the internal magnetic field are calculated in Sec. 3.1.

\begin{figure*}
 \includegraphics[width=4cm]{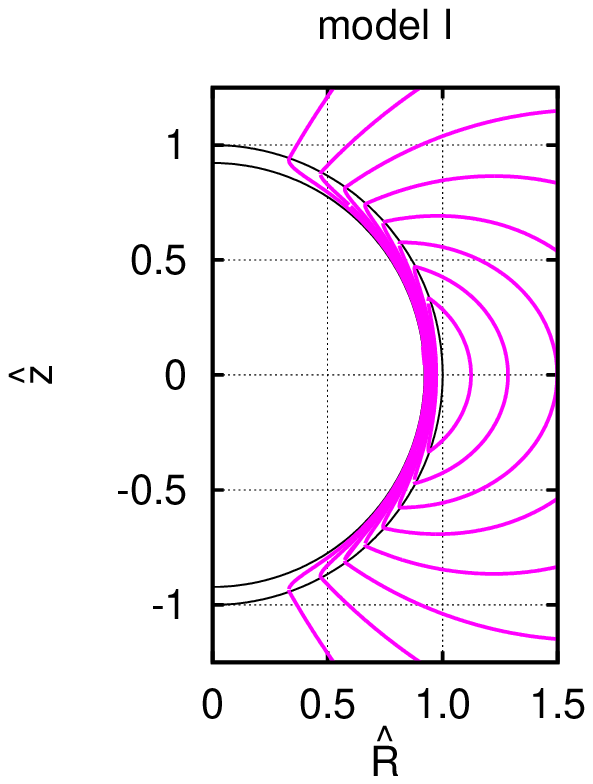}
 \includegraphics[width=4cm]{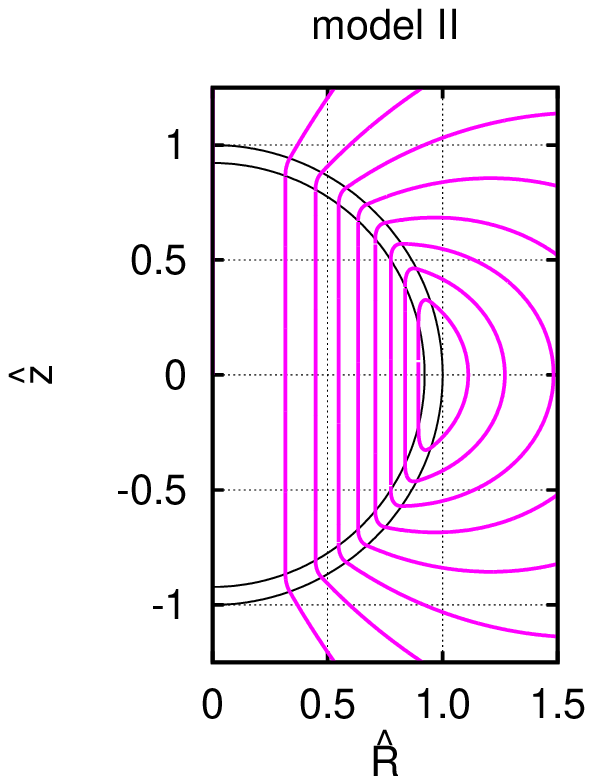}
 \includegraphics[width=4cm]{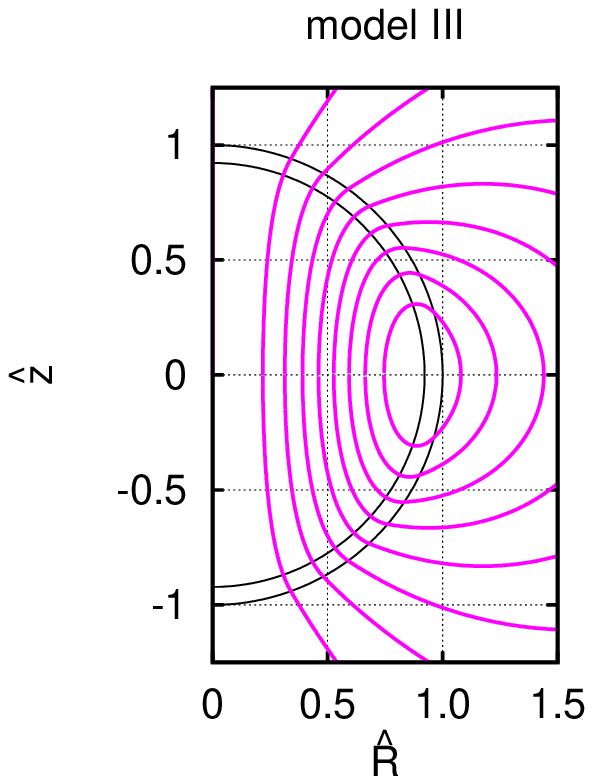}
 \includegraphics[width=4cm]{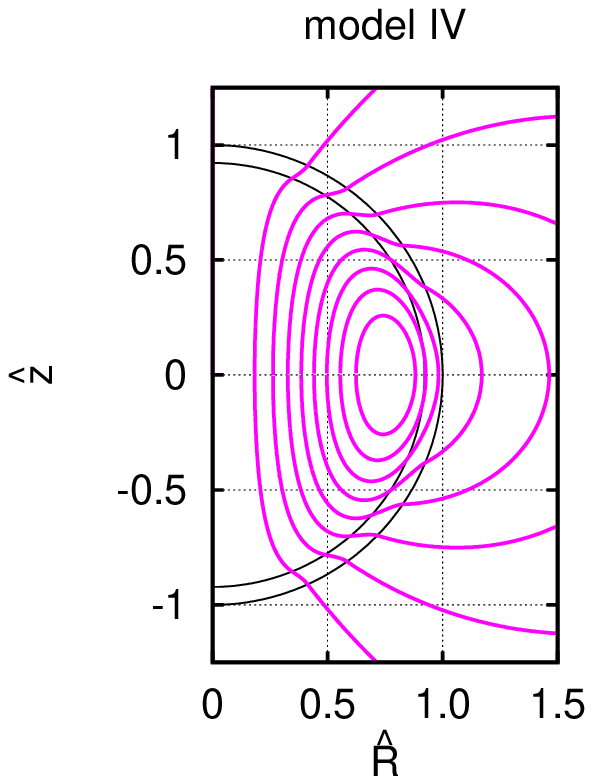}
\caption{The contours of $\hat{\Psi}$ in each analytical model (see App. \ref{App:analytical}). 
The inner curve is the core-crust boundary
and the outer curve is the stellar surface.
Model I: This model has purely crustal open magnetic fields.
There is a negative current sheet on the core-crust boundary in order to 
exclude the core magnetic fields. 
 Model II: This model has a purely crustal toroidal current. The
core magnetic field is an inner vacuum solution of the
crustal toroidal current.
Model III: This model has both crustal and core toroidal currents.
The configuration of the core magnetic field is different from Model II.
Model IV: This model has opposite flowing toroidal current density.
The core magnetic fields are stronger than crustal magnetic fields.
The $\hat{z}$ and $\hat{R}$ denote dimensionless forms of $z$ and $R$.
We will define the dimensionless forms in Sec. 2.4.
}
 \label{Fig:models}
\end{figure*}

In order to achieve the core-crust conditions described, 
we choose the simplest forms (\citealt{Gourgouliatos_et_al_2013})
\begin{eqnarray}
 S(\Psi) = S_0,
\end{eqnarray}
\begin{eqnarray}
 F(\Psi)  = F_0.
\end{eqnarray}
Although we can also compute
more complex functional forms
(see \citealt{Fujisawa_Yoshida_Eriguchi_2012}; \citealt{Ciolfi_Rezzolla_2013}),
we used the simplest functional forms.
We are interested in magnetic field configurations 
throughout the star and the influence of the boundary conditions.
These functional forms were used to examine them easily and clearly. 
In particular, the values of $S_0$ and $F_0$ satisfy the conditions of $S_0 F_0 > 0$ (model III) and
$S_0 F_0 < 0$ (model IV) in actual numerical computations.

We also chose the functional form of $I$
to satisfy the boundary conditions on 
the stellar surface and the core-crust surface. 
If we assume that the magnetosphere is not twisted ($B_\varphi = 0$, outside the star), 
the toroidal current density must vanish outside the star ($I = 0$).
We need to choose a functional form, such as the following (\citealt{Tomimura_Eriguchi_2005})
\begin{eqnarray}
 I(\Psi) = 
  \left\{ 
\begin{array}{lc}
I_0(\Psi - \Psi_{\rm{ex}, \max})^{k+1} & (\Psi > \Psi_{\rm{ex}, \max})  \\
0 & (\Psi \leq \Psi_{\rm{ex}, \max})
\label{Eq:I}
\end{array}
\right.
,
\end{eqnarray}
where $\Psi_{\rm{ex}, \max}$ is the maximum value of $\Psi$ in the stellar exterior.
We fixed the parameter $k=0.1$ in all of our numerical computations. 
This choice of the parameter is exactly the same as in \cite{Yoshida_Eriguchi_2006} 
(also \citealt{Yoshida_Yoshida_Eriguchi_2006};
\citealt{Lander_Jones_2009}; \citealt{Ciolfi_et_al_2009};
\citealt{Fujisawa_Yoshida_Eriguchi_2012}). 
As many previous works found, the choice of $k = 0.1$ results in
locally strong toroidal magnetic fields.

The functional form of Eq. (\ref{Eq:I}) is available in models I, III, and IV. 
When we calculated model II, we set $\Psi_{\rm{ex},\max}$ as the maximum value of
$\Psi_{\rm{ex},\max}$ and $\Psi_{\rm{c},\max}$, where $\Psi_{\rm{c},\max}$ is the maximum value of $\Psi$ 
in the core. If we choose this functional form, the current density vanishes outside the star
and there is no toroidal magnetic field. This functional form always satisfies the boundary condition
$I=0$ at the stellar surface.

\subsection{Models of a twisted magnetosphere}

We calculated two types of twisted 
force-free magnetospheres:  
the equatorial shearing model and the ring  model
(see \citealt{Parfrey_Beloborodov_Hui_2013}).
To obtain the equatorial sharing model,  we  chose the following functional form:
\begin{eqnarray}
 I(\Psi) = 
  \left\{ 
\begin{array}{lc}
I_0(\Psi - \Psi_{t, \min})^{k+1} & (\Psi > \Psi_{t, \min})  \\
0 & (\Psi \leq \Psi_{t, \min})
\end{array}
\right.
.
\label{Eq:I1}
\end{eqnarray}
For the ring shearing model, we chose the following functional form:
{\scriptsize
\begin{eqnarray}
I(\Psi) = 
 \left\{ 
\begin{array}{ll}
I_0(\Psi - \Psi_{\rm{ex}, \max})^{k+1} & (\Psi > \Psi_{\rm{ex}, \max})  \\
0 & (\epsilon \Psi_{\rm{ex},\max} \leq  \Psi \leq \Psi_{\rm{ex},\max}) \\
I_1 \{(\epsilon \Psi_{\rm{ex},\max} - \Psi) (\Psi - \Psi_{t,\min})\}^{k_2} 
&(\Psi_{t,\min} \leq \Psi \leq \epsilon \Psi_{\rm{ex},\max}) \\
0 & (\Psi < \Psi_{t,\min})
\end{array}
\right.
,
\label{Eq:I2}
\end{eqnarray}
}
Here, 
$I_1$, $\epsilon$ and $k_2$ are the parameters of a twisted magnetosphere.
$\Psi_{t, \min}$ denotes the minimum value of $\Psi$ within the 
twisted magnetosphere.
We defined the value of the maximum radius of the twisted field as 
$r_M$. The twisted field in the magnetosphere cannot exist beyond $r_M$.
The regions of the twisted magnetosphere are limited by the 
last closed magnetic field line inside the radius $r_M$ using these functional forms.
When the twisted field in the magnetosphere is the ring model, the twisted field can exist 
within the limited region between the field 
lines $\Psi = \Psi_{t,\max}$ and $\Psi = \epsilon \Psi_{\rm{ex},\max}$ (see Eq. \ref{Eq:I2}).
The value of $\epsilon$ is bounded by $0 < \epsilon < 1$. This parameter
determines the width of the twisted field region.
We fixed $k_2 = 1.0$ and $\epsilon = 0.5$ in the ring model.
We examined $k_2 > 1.0$ models, but did not obtain toroidal magnetic field solutions as
strong as for $k_2= 1.0$. As a result, we found  that $k_2 = 1.0$ results in 
locally strong twisted magnetic fields within the ring regions.

\subsection{Equation of state and numerical setting}

We introduce a realistic EOS of neutron stars and a numerical setting.
\cite{Glampedakis_Lander_Andersson_2014} used an $N=1$ polytrope  
as a simple example of a neutron stars' EOS. They were not able to
distinguish between the core and crust regions due to the simple polytropic EOS.
Note that the magnetic field configurations in the equilibrium state 
depend on the distributions of mass density in the core and the 
electron number density within the crust (Eqs. \ref{Eq:jphi_Hall} and \ref{Eq:jphi_MHD}). 
Therefore, realistic values of the size and 
the composition of the crust are required to calculate 
the equilibrium state of the crust and the core. We employed
SLy EOS (\citealt{Douchin_Haensel_2001})
for our EOS. Since SLy EOS describes both the crust and the liquid core,
it is suitable for the magnetar interior.
This EOS can describe $\sim 2 M_\odot$ neutron stars (\citealt{Antoniadis_et_al_2013}; 
\citealt{Demorest_et_al_2010}).

\begin{figure*}

\begin{center}
 \includegraphics[width=7cm]{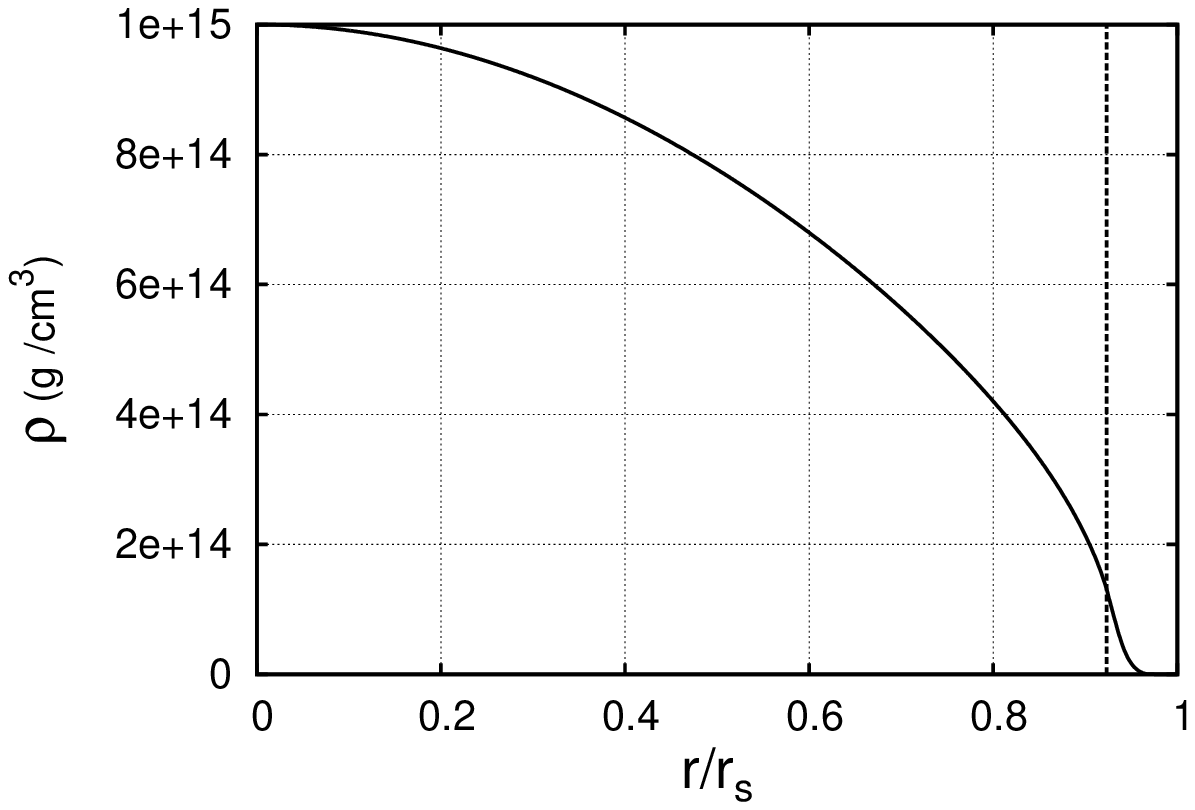}
 \includegraphics[width=7cm]{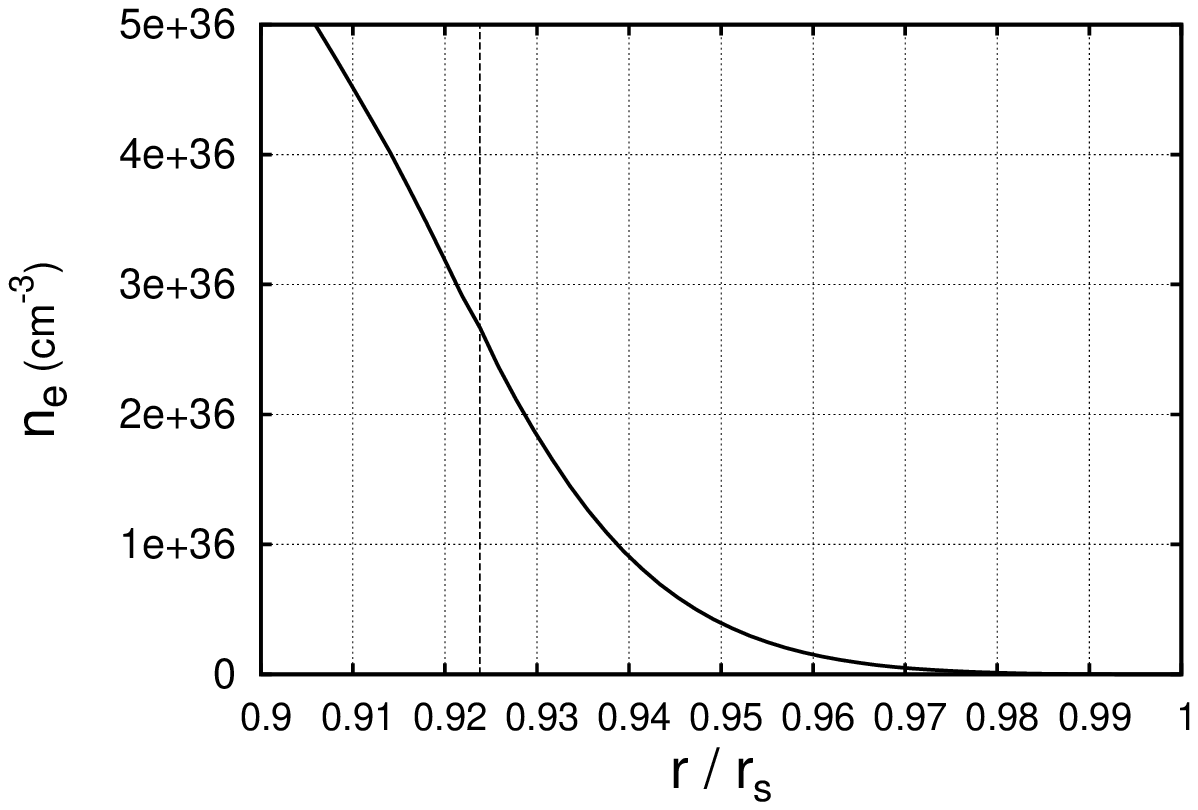}
\end{center}

\caption{
Mass density profile (left) and electron number density profile
 near the crust (right). 
The horizontal axis is normalized by the stellar radius $r_s$.
Dashed lines correspond to the crust-core interface at $ r_{\rm in}/r_s \sim 0.923$.
}
\label{Fig:rho_n_e}
\end{figure*}

We fixed the maximum density $\rho_{\max} = 1.0 \times 10^{15} \mathrm{g/cm^{3}}$ 
and obtained the distributions of the mass density and electron number density (Fig. \ref{Fig:rho_n_e})
using SLy EOS. 
The obtained stellar radius is $r_s \sim 1.17 \times 10^{6} \rm{cm}$ and 
the mass is $\sim 1.42 M_\odot$.
The radius of the core-crust interface $r_{\rm in}$ is $r_{\rm in} / r_{s} \sim 0.923$
in this model,  such that the thickness of the crust is approximately $\sim 8 \%$ of the stellar radius.
The electron number density at the 
base of the crust is $n_e \sim 2.67 \times 10^{36} \mathrm{cm^{-3}}$ in this model. 
We defined $n_c \equiv 2.67 \times 10^{36} \mathrm{cm^{-3}}$ as the maximum electron number 
density in our formulation.
Noted that we assume the star is spherical in this paper, 
because the deformation due to magnetic fields is small, even a  magnetar
(\citealt{Haskell_et_al_2008}; \citealt{Ciolfi_Rezzolla_2013}).
Therefore, we used this spherical magnetar model in 
our numerical computations.

To perform numerical computations properly,
 we introduced the  following dimensionless physical quantities:
\begin{eqnarray}
 \hat{r} = \frac{r}{r_s}, \hspace{10pt} \hat{R} = \frac{R}{r_s}, \hspace{10pt} \hat{z} = \frac{z}{r_s},
\end{eqnarray}
\begin{eqnarray}
\hat{\rho} = \frac{\rho}{\rho_{\max}},
\end{eqnarray}
\begin{eqnarray}
 \hat{n}_e = \frac{n_e}{n_c}.
\end{eqnarray}
We need one more physical quantity
in order to define dimensionless forms of other physical quantities.
We chose $S_{\max}$ (the maximum value of $S$) as the physical quantity and
defined the dimensionless $\hat{S}$ as,
\begin{eqnarray}
 \hat{S} = \frac{S}{|S_{\max}|}.
\label{Eq:hat_S}
\end{eqnarray}
Using these four quantities, we can obtain the dimensionless 
magnetic energy: 
\begin{eqnarray}
 \hat{E} = \frac{E}{r_s^5 n_{c}^2 S_{\max}^2}.
\label{Eq:hat_E}
\end{eqnarray}
We also show the forms 
of the other quantities in App. \ref{App:dimensionless}.
By using these quantities, the functional forms of arbitrary 
functions also  become dimensionless. Note that the value of $\hat{S}_0$ is $1$ or $-1$ 
because of the functional form $S(\Psi) = S_0$ and the definition in Eq. (\ref{Eq:hat_S}).

In order to evaluate the energy ratio, we calculated the poloidal 
magnetic energy $\hat{E}_p$ and the toroidal magnetic energy $\hat{E}_t$ as
\begin{eqnarray}
 \hat{E}_p = \frac{1}{8\pi} \int ( \hat{B}_r^2 + \hat{B}_\theta^2 ) d\hat{V},
\end{eqnarray}
\begin{eqnarray}
 \hat{E}_t = \frac{1}{8\pi} \int \hat{B}_\varphi^2  d\hat{V}.
\end{eqnarray}
We defined the total magnetic energy  $\hat{E} = \hat{E}_p + \hat{E}_t$.
We also calculated the core magnetic energy ($\hat{E}\rm{co}$), the crust magnetic energy ($\hat{E}\rm{cr}$),
and the stellar interior magnetic energy ($\hat{E}\rm{st} = \hat{E}{co} + \hat{E}\rm{cr}$).
Our numerical domain is defined as $0 \leq \theta \leq \pi$  in the 
angular direction  with the stellar interior ($0 \leq \hat{r} \leq 1$) 
exterior magnetosphere ($1 \leq \hat{r} \leq 2$ for non-twisted magnetosphere models or
$1 \leq \hat{r} \leq 16$ for twisted magnetosphere models) in the radial direction.
We  used a sufficient number of mesh  points in this paper 
(see the analytical solutions in App. \ref{App:analytical}
and accuracy verification in App. \ref{App:accuracy}).

\section{Results}
 \label{Sec:Numerical_result}

\subsection{Core-crust solutions with a non-twisted magnetosphere}
\label{Sec:core_magnetic}

\begin{figure*}
\begin{center}
\includegraphics[width=5.2cm]{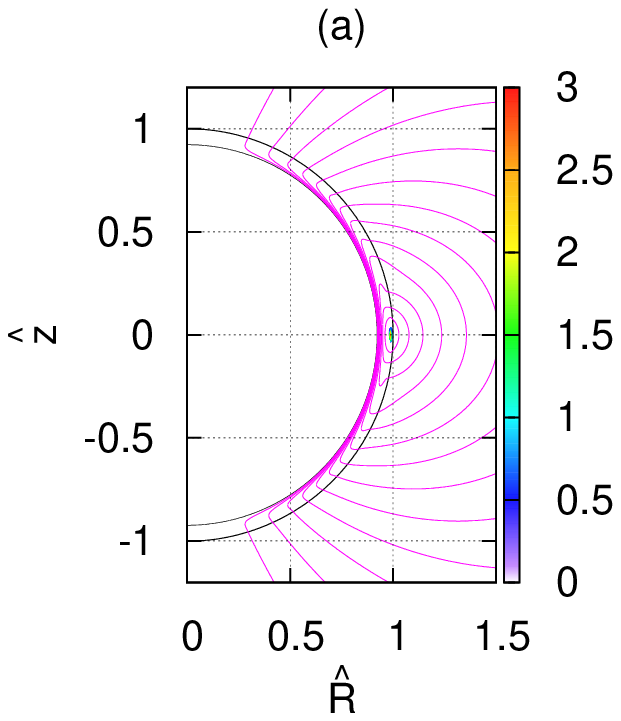}
\includegraphics[width=5.2cm]{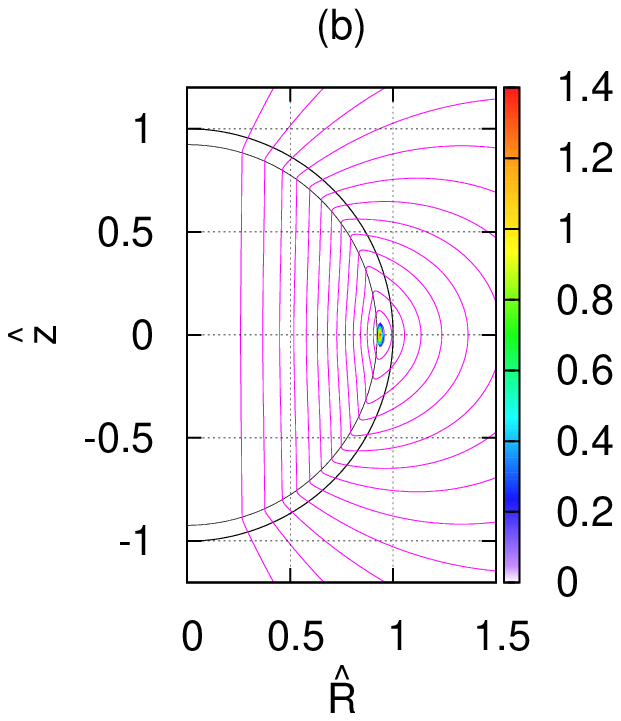}
\includegraphics[width=5.2cm]{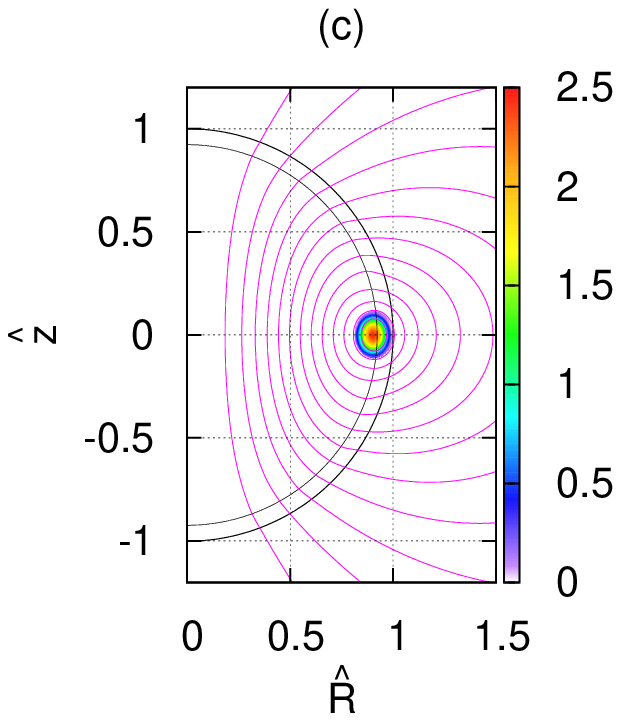}

\includegraphics[width=5.2cm]{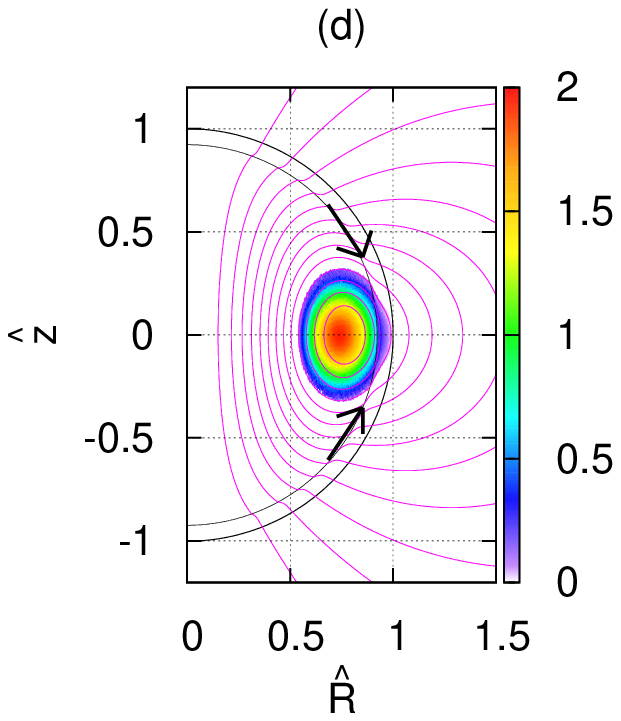}
\includegraphics[width=5.2cm]{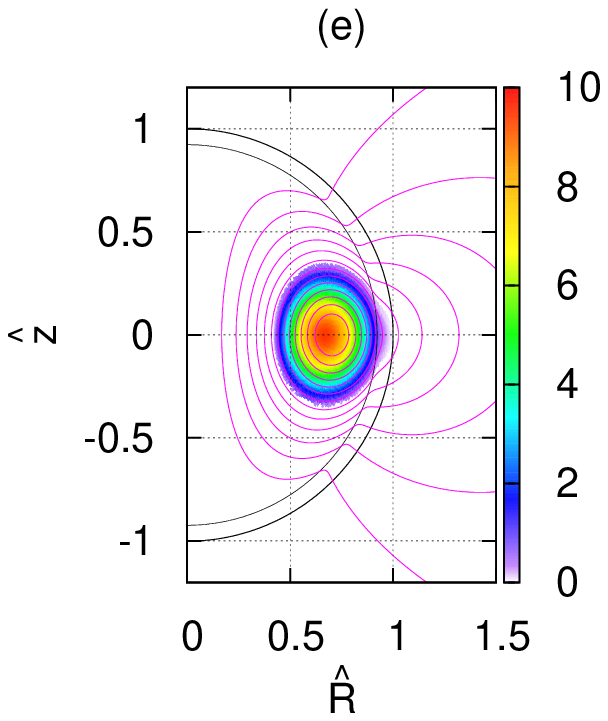}
\includegraphics[width=5.2cm]{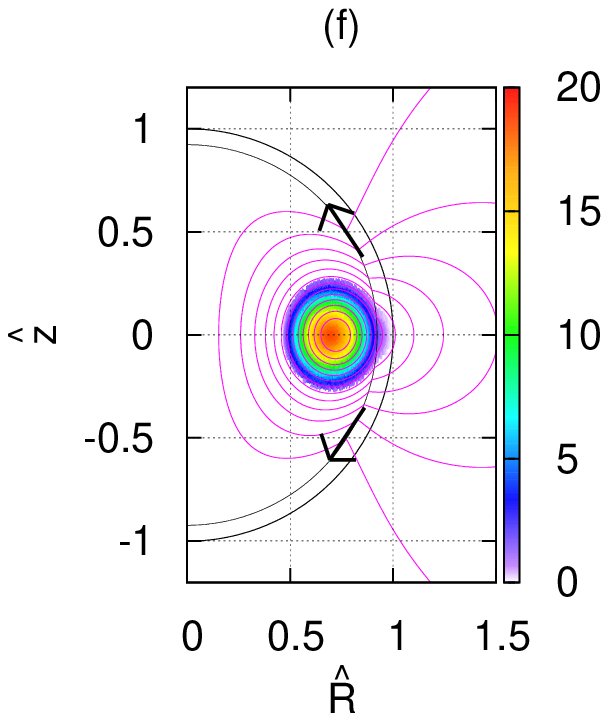}
\end{center}

\caption{The contours of $\hat{\Psi}$ of each model (solid magenta lines). 
The black inner curve and the outer curve denote
the core-crust boundary and the stellar surface, respectively. The colour maps denote 
the magnitude of the toroidal magnetic field $\hat{B}_\varphi$ normalized by
 the strength of the dipole component at the north pole ($\hat{B}_d$).
The arrows denote the directions of magnetic stress due to the current sheet.}
\label{Fig:fig1}
\end{figure*}

\begin{table*}
 \begin{tabular}{ccccccccccccc}
\hline
  & $\hat{I}_0 $ & $\hat{S}_0 $ & $\hat{F}_0 $ & model & $E{\rm cr}_t / E{\rm cr}$ & $E{\rm co}_t / E{\rm co}$ 
&  $E{\rm cr} / E$ & $E{\rm co} / E$ & $\hat{E}$& $\hat{B}_d $&  $\hat{j}_0$ \\
\hline
  (a)  &  800  & 1     & 0   & I   & 1.23E-4  & 0.0     & 0.98 & 0.00 & 5.99E-5 & 3.55E-3 & - \\ 
  (b)  &  250  & 1     & 0   & II  & 1.40E-2  & 0.0     & 0.08 & 0.67 & 3.27E-3 & 1.06E-1 & - \\
  (c)  &   30  & 1     & 1   & III & 1.14E-1  & 2.41E-2 & 0.07 & 0.74 & 3.27E-1 & 8.75E-1 & 0.0 \\
  (d)  &   10  &-1     & 0.1   & IV  & 4.09E-3  & 6.11E-2 & 0.08 & 0.83 & 1.88E-3 & 4.71E-2 & $0.1$ \\
  (e)  &   10  &-1     & 0.1   & IV  & 7.73E-3  & 2.68E-1 & 0.02 & 0.95 & 9.83E-3 & 4.75E-2 & 0.0 \\
  (f)  &   10  &-1     & 0.1   & IV  & 1.62E-2  & 3.37E-1 & 0.02 & 0.96 & 8.81E-1 & 3.01E-1 & $-1.0$ \\
\hline

 \end{tabular}
\caption{Parameters and numerical solutions of the models.
$\hat{j}_0$ denotes the strength of the current sheet on the core-crust interface.
$\hat{B}_d$ is the dimensionless strength of the dipole magnetic field.}
\label{Tab:tab1}
\end{table*}

First, we will show the solutions 
with a non-twisted magnetosphere ($\hat{B}_\varphi = 0$, outside the star). 
Since the toroidal magnetic field energy ratio
(within the crust, $E{\rm{cr}}_t  / E{\rm{cr}}$) 
relates to the Hall drift activity within the crust (\citealt{Kojima_Kisaka_2012};
\citealt{Vigano_et_al_2013}),
it is of  interest to characterize the magnetized equilibrium states.
We will focus on the energy ratios $E{\rm{cr}}_t / E{\rm{cr}}$ in this paper. 
We computed many equilibria,  changing the
value of $\hat{I}_0$ for the 6 solutions types 
mentioned in Sec. 2.2. Each model displayed here 
has the largest value of the energy ratio, $E{\rm cr}_t / E{\rm cr}$, 
for the present functional forms.
We labelled the 6 obtained solutions (a) -- (f) as follows:
solution (a) model I type, solution (b) model II type, 
solution (c) model III type, solution (d) model IV type with a positive current sheet,
solution (e) model IV type, solution (f) model IV type with a negative current sheet. 
The numerical solutions are tabulated in Tab. \ref{Tab:tab1}.
The configurations of the poloidal
and the toroidal magnetic fields are displayed in Fig. \ref{Fig:fig1}.
The magnitude of the toroidal magnetic fields in Fig. \ref{Fig:fig1} is
normalized by that of the  magnetic dipole component of each solution 
at the stellar north pole. As seen in Tab. \ref{Tab:tab1}, 
the energy ratio of the core ($E{\rm co}/E$) increases  sequentially
from solution (a) to solution (f).
The value reaches $E{\rm co} / E \sim 0.96$ in solution (f),
while $Eco / E \sim 0.67$ in solution (b).
Since solutions (d) and (f) have current sheets on the core-crust boundary, 
the poloidal field lines bend at this boundary.
The direction of the bending depends 
on the sign of the current sheet.
The direction of the magnetic 
stress due to the current sheet is plotted in Fig. \ref{Fig:fig1} .

The energy ratio$E{\rm cr}_{t}/E{\rm {cr}}$  for each of the models is smaller than 0.5,
such that the energy of poloidal components is dominant.
However, the energy ratio changes according to the models.
As seen in Tab. \ref{Tab:tab1} and  Fig. \ref{Fig:fig1}, 
the region of the toroidal magnetic field of solution (a) is very small, as is
the energy ratio ($E{\rm cr}_t/E{\rm cr} \sim 10^{-4}$).
This value is much smaller than in the solution by 
\cite{Gourgouliatos_et_al_2013} because the width of the crust derived from a realistic EOS is 
smaller (see Fig. 9 in their paper).
The energy ratio $E{\rm cr}_t / E{\rm cr}$ of solution (b) is larger than that of solution (a).
The toroidal magnetic field region also becomes 
slightly larger.
Since the poloidal magnetic field
lines penetrate the core region in this model,
the $\hat{\Psi} \geq  \hat{\Psi}_{\max}$ region within the crust
becomes large (compare the analytical profiles of model I with model II in Fig. \ref{Fig:models_r}).

Solution (c) is a model III type solution
with core toroidal current density.
The energy ratio of the core  ($E{\rm co} / E$) of
solution (c) is larger than that of solution (b).
The energy ratio ($E{\rm cr}_t / E{\rm cr}$) of solution (c)
reaches 0.11 and is larger than those of 
solutions (a) and (b).
On the other hand, the core toroidal magnetic fields 
are almost zero in solution (c).
This numerical result shows that 
the core magnetic field structure is important 
when  considering $E{\rm cr}_t / E{\rm cr}$ in Hall equilibrium within the crust.

Solutions (d),  (e), and (f) are model IV type solutions. 
They have opposite toroidal 
currents because the value of $\hat{F}_0 \hat{S}_0$ is always negative.
As seen in Tab. \ref{Tab:tab1}, 
such oppositely flowing crustal toroidal current increases the core energy ratios 
$(E{\rm co}/E \& E{\rm co}_t / E{\rm co}$). 
These tendencies are consistent with the arguments by \cite{Fujisawa_Eriguchi_2013}.
The ratios $E{\rm co}/E$ reach approximately 0.8 -- 0.9. 
These values are the largest among our solutions. 
These solutions also sustain the large toroidal magnetic field energy 
in the core region.
The energy ratios $(E{\rm co}_t / E{\rm co})$ are  $\sim 0.2-0.3$.
The size of the toroidal magnetic field region within the crust of model IV solutions 
 is slightly larger than those of any other solutions. 
The boundary conditions of $\hat{I}$ at the core-crust boundary are
$\hat{I} \neq 0$ and the values of $\hat{I}$ continue smoothly at the boundary.
These numerical results indicate that the $\hat{I} \neq 0$ boundary conditions
on the core-crust boundary broaden the size of the 
toroidal magnetic field regions within the crust of the Hall equilibrium.

Solutions (d) and (f) have positive and negative current sheets 
 on the core-crust boundary, respectively.
 Although the physical meaning and origin of the current sheet are unclear and difficult 
to understand  in terms of micro physics (\citealt{Lander_2013a, Lander_2014}), 
we can regard the current sheet under the magnetic fields as 
magnetic stress on the bottom of the crust (\citealt{Braithwaite_Spruit_2006}). 
In this interpretation, the sign of the current sheet indicates the direction of the stress.
The positive and negative current sheets denote
the magnetic stresses in the direction of the equator and poles, respectively 
(see arrows in Fig. \ref{Fig:fig1}). 
As seen in Tab. \ref{Tab:tab1}, a positive current sheet decreases the
energy ratio $E{\rm co} /E$ (see solutions c and d), while
a negative current sheet increases the ratio $E{\rm co} / E$ (see solutions (e) and (f) ).
Moreover, solution (d) has a smaller value of $E{\rm cr}_t /Eco$ 
than solution (e). 
Solution (f) also has larger values of $E{\rm cr}_t / E$ and $E{\rm co}_t /E$ than solution (e).
Note that the maximum magnitude of the core toroidal magnetic 
fields in solution (f) reaches approximately 20 times the size of
its dipole magnetic component at the north pole (Fig. \ref{Fig:fig1}).
This means that the current sheet (magnetic stress) 
at the bottom of the crust can sustain the strong 
core toroidal magnetic fields (see also Sec. 6.4 in \citealt{Fujisawa_Eriguchi_2013}).
These results show that the core-crust stress affects the energy ratios of
both crustal and core toroidal magnetic fields. 

The energy ratio $E{\rm cr}_t / {\rm E}$
is affected by the boundary conditions
and the core magnetic field configurations. 
This means that the Hall MHD evolution 
is strongly affected by core magnetic field conditions
as per the  recent numerical simulation by \cite{Vigano_et_al_2013}.
We will discuss  the influence of the boundary  condition on the Hall MHD evolution
in Sec. \ref{subsec:MHD_evolution}.

\subsection{Solutions with  an equatorial shearing magnetosphere}
\label{Sec:equatorial_shearing}

\begin{table*}
 \begin{tabular}{cccccccccccccc}
  \hline
&  $\hat{r}_M$ & $\hat{I}_0$ & $\hat{S}_0$ & $\hat{F}_0$ & model &  
 $E{\rm cr} / E$ & $E{\rm co} / E$ &   $E{\rm st}/E$ & $\hat{E}$ & $\hat{B}_d$ & $\hat{r}_x$\\
\hline
(m-a) &  8.0   &   0.1  & 1.0     & 0.0   & I   & 0.98 & 0.0 & 0.98  & 5.89E-5 & 3.41E-3 & - \\
(m-b) &  8.0   &   1.5  & 1.0     & 0.0   & I   & 0.98 & 0.0 & 0.98  & 5.93E-5 & 3.47E-3 & - \\
(m-c) &  8.0   &   2.0  & 1.0     & 0.0   & I   & 0.98 & 0.0 & 0.98  & 5.94E-5 & 3.48E-3 & 3.16 \\
\hline
(m-d) &  8.0   &   0.1  & -1.0    & 0.5   & IV   & 0.03 & 0.89 & 0.92  & 2.17E-2 & 1.47E-1 &   -   \\
(m-e) &  8.0   &   1.1  & -1.0    & 0.5   & IV   & 0.03 & 0.86 & 0.89  & 2.58E-2 & 1.70E-1 &   -   \\
(m-f) &  8.0   &   1.2  & -1.0    & 0.5   & IV   & 0.03 & 0.85 & 0.88  & 2.65E-2 & 1.73E-1 & 3.29  \\
 \hline
 \end{tabular}
\caption{Parameters and solutions with equatorial shearing models. Model I magnetic fields (m-a, m-b, m-c)
and model IV magnetic fields (m-d, m-e, m-f) are tabulated.}
 \label{Tab:m-e}
\end{table*}

\begin{figure*}
\begin{center}
 \includegraphics[width=5.2cm]{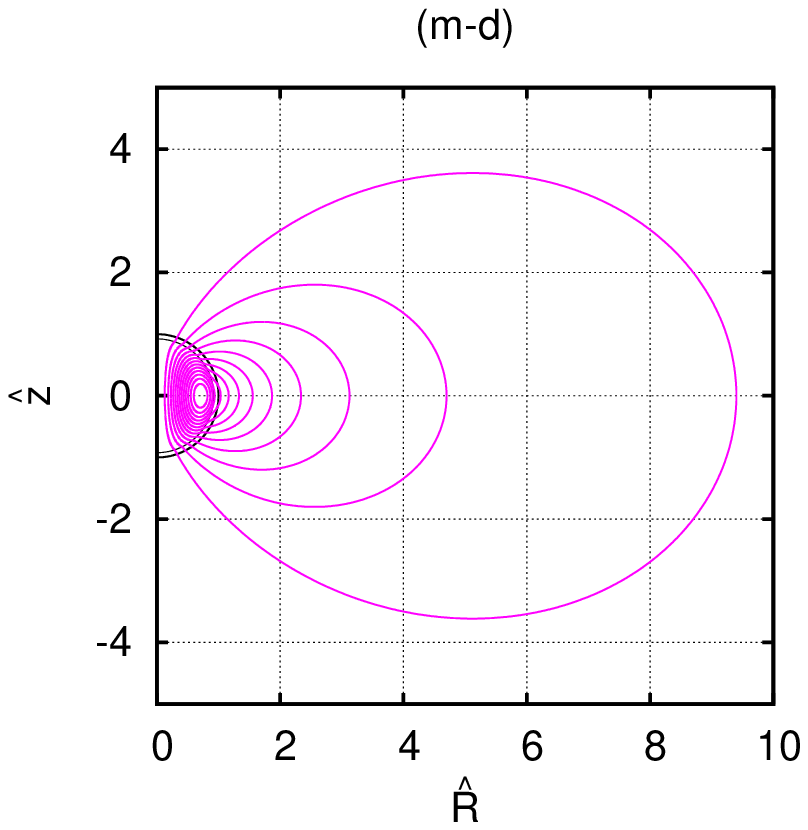}
 \includegraphics[width=5.2cm]{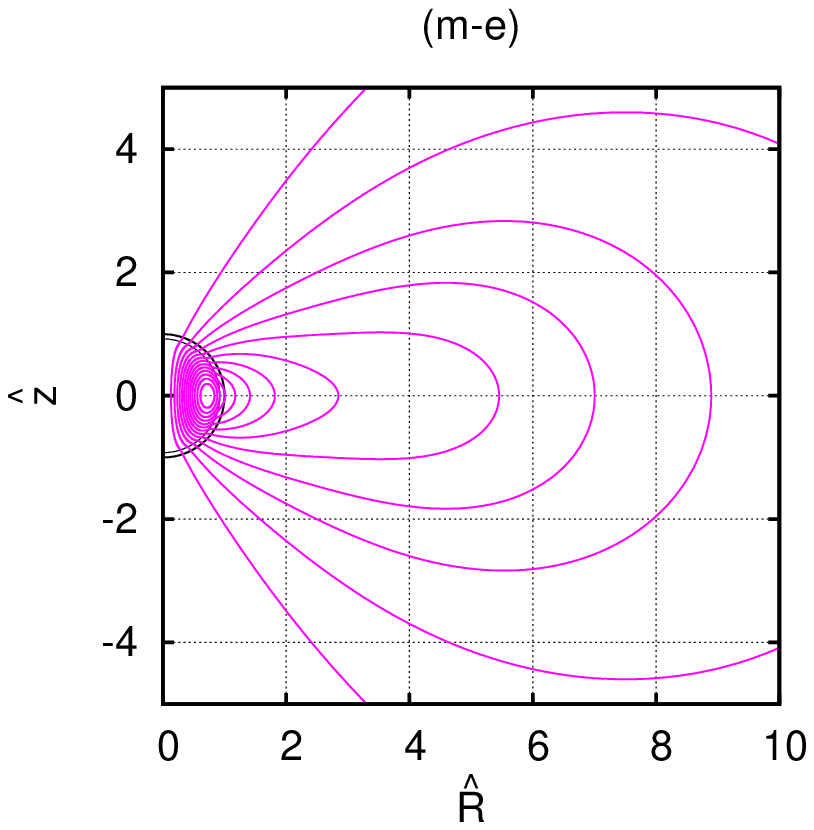}
 \includegraphics[width=5.2cm]{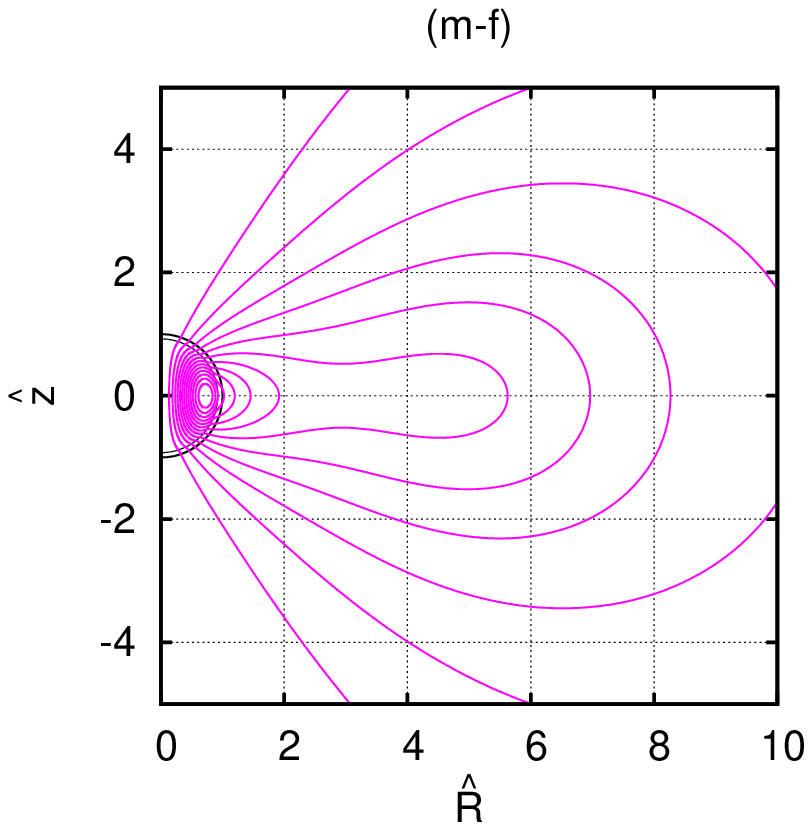}
 
 \includegraphics[width=5.2cm]{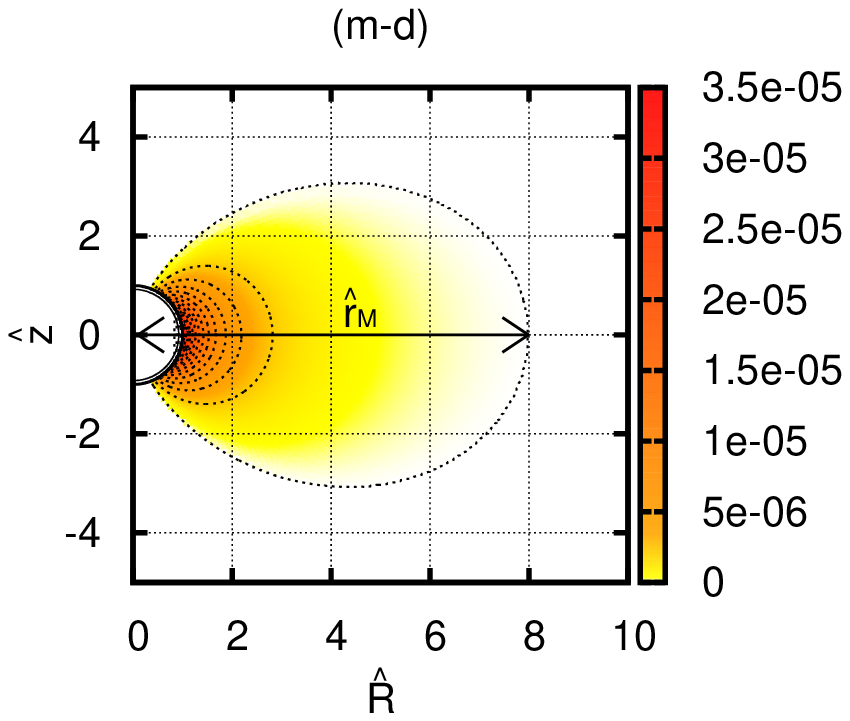}
 \includegraphics[width=5.2cm]{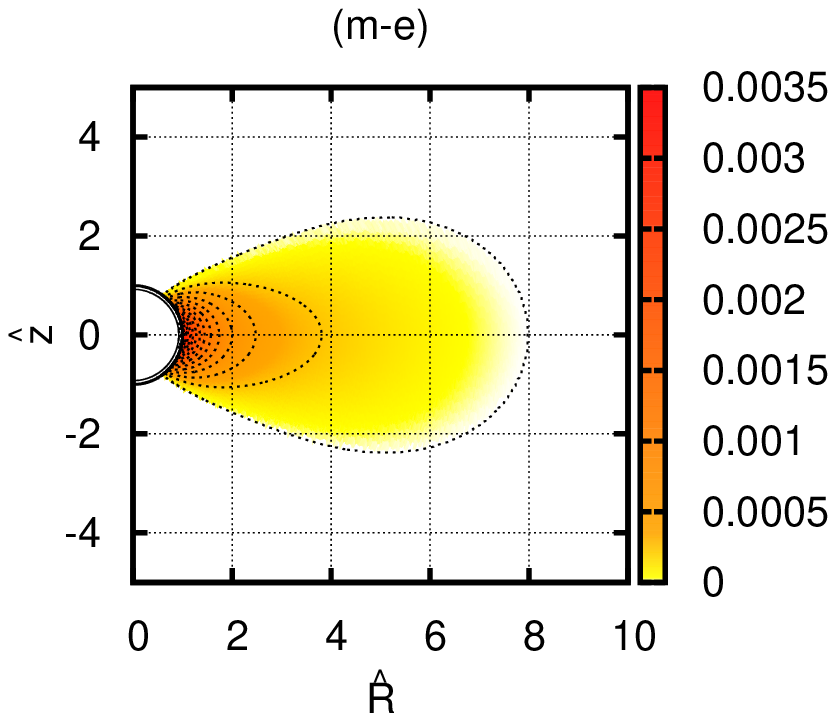}
 \includegraphics[width=5.2cm]{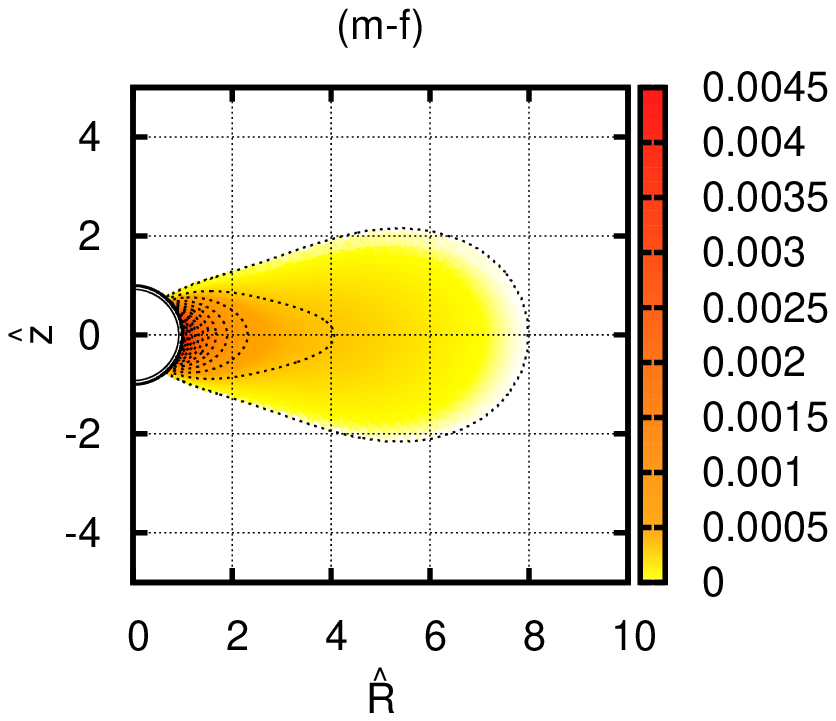}
\end{center}
\caption{Top panels: The contours of $\hat{\Psi}$.
Bottom panels: The colour maps and contours of $\hat{j}_\varphi$ in a twisted magnetosphere.
 The inner circles in all panels denote core-crust interfaces and stellar surfaces.  
 The X-point geometry appears at $\hat{r} \sim 3.29$ on the equatorial plane in solution (m-f) (top right panel).
The arrow in the bottom left panel denotes the size of $\hat{r}_M$ .
}
\label{Fig:m}
\end{figure*}

\begin{figure*}
  \begin{center}
   \includegraphics[width=8cm]{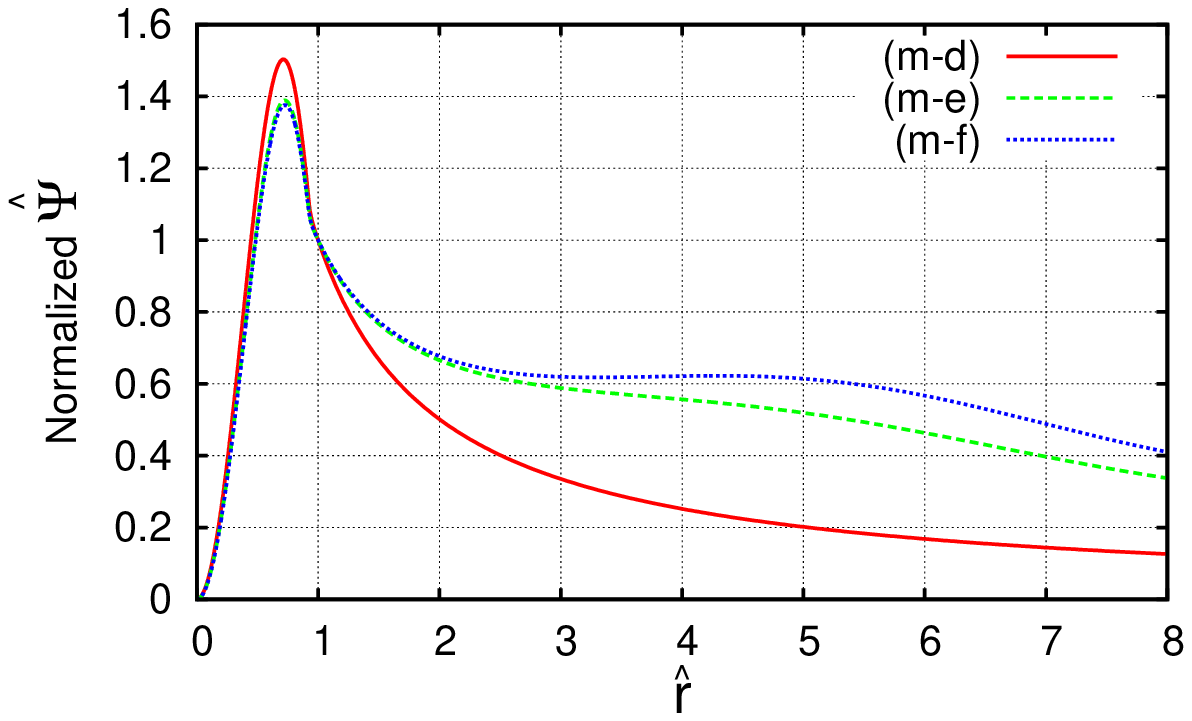}
   \includegraphics[width=8cm]{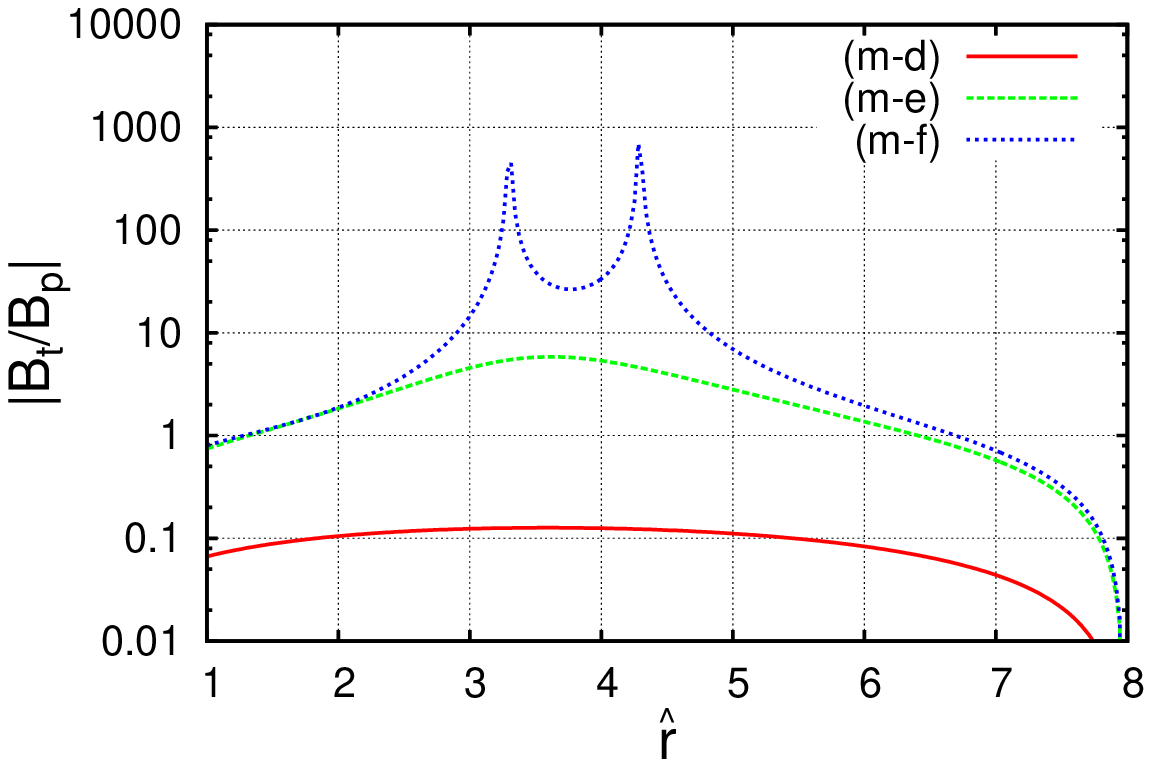}
 \end{center}
 \caption{
Left panel : The profiles of $\Psi$ normalized by the surface value on the equatorial plane.
 Each line denotes solution (m-d) (solid line), solution (m-e) (dashed line), and solution (m-f) (dotted line). 
 Solution (m-d) and solution (m-e) decrease monotonically in the magnetosphere ($\hat{r} > 1$), 
 but solution (m-f) has a local maximum value at $\hat{r} \sim 3.29$. 
 Right panel : The ratios of the toroidal component and the poloidal component $|B_t / B_p|$. 
 These values represent the strength of the twist of the fields.
 Solution (m-f) has two singular points
at $\hat{r} \sim 3.29$ and $\hat{r} \sim 4.2$. The first singular point comes from the X-point
geometry and the second point comes from the magnetic neutral point (see Fig. \ref{Fig:m})}
\label{Fig:BtBp}
\end{figure*}

We calculated solutions with 
equatorial shearing using the functional form $\hat{I}(\hat{\Psi})$ in Eq. (\ref{Eq:I1}).
First,  we will show six solutions by changing the value of $\hat{I}_0$. 
We used two types of internal magnetic field models (I, IV).
Solutions (m-a), (m-b) and (m-c)  have model I internal fields and
solutions (m-d), (m-e) and (m-f) have model IV internal fields.
Solutions (m-d), (m-e) and (m-f) do not have current sheets on the 
core-crust boundaries. 
 The numerical results are tabulated in Tab. \ref{Tab:m-e} and
 are displayed in Figs. \ref{Fig:m} and \ref{Fig:BtBp}.
 Three contours of $\hat{\Psi}$ -- (m-d), (m-e), and (m-f) -- and magnetospheric toroidal 
current distributions are displayed in Fig. \ref{Fig:m}. 
The arrow in the bottom left panel in Fig. \ref{Fig:m} denotes the
size of $\hat{r}_M$.
The profiles of  $\hat{\Psi}$ and the ratio of $B_t$ to $B_p$
on the equatorial plane are also displayed in Fig. \ref{Fig:BtBp}. 
The equatorial shearing becomes larger
as the value of $\hat{I}_0$ increases because the value of 
$\hat{I}_0$ represents the twisted strength (toroidal component) 
of the magnetic fields.

As seen in Fig. \ref{Fig:m}, the configuration of the poloidal magnetic field in the 
magnetosphere
changes as shearing (the value of $\hat{I}_0$) increased. 
The poloidal magnetic field lines near the equatorial plane 
are stretched outward by the equatorial shearing current (compare (m-d) with 
(m-e) in Fig. \ref{Fig:m}).  
Since the energy ratio of stellar magnetic energy to the total magnetic energy $E{\rm st}/E$ 
decreases as the value of $\hat{I}_0$ increases 
from Tab. \ref{Tab:m-e}, the value $E - E{\rm st}$ increases.
 We can see an interesting magnetic field structure near 
$\hat{r} \sim 3.3$ in Fig. \ref{Fig:BtBp}} (solution (m-f)). 
As seen in the left panel, $\hat{\Psi}$ of solutions (m-d) and (m-e) decrease monotonically 
outside the star ($\hat{r} \geq 1$),  but solution (m-f) has a local minimum
near $\hat{r} \sim 3.3$ where the $\hat{r}$ derivative of $\hat{\Psi}$ becomes $0$.
Since the sign of $\partial \hat{\Psi} / \partial \hat{r}$
represents the direction of $\hat{B}_\theta$, the direction of the poloidal magnetic 
fields reverses at this point.
Thus, the poloidal magnetic field lines cross 
and an X-point geometry forms at the point.
We see X-points as singularities in Fig. \ref{Fig:BtBp}.
As the value of $\hat{I}_0$ increases, the magnitude of the twist ($|B_t / B_p|$)
also increases (from (m-d) to (m-f) in the panels).
If the twist reaches a critical value, the poloidal magnetic fields vanish and 
an X-point geometry forms. The inner singular point represents
X-point geometry and the outer one represents the 
centre of the magnetic loop (see also solution (m-f) in Fig. \ref{Fig:m}). 
We define $\hat{r}_X$ as the distance from the centre to the inner singular point.

We can  also find X-point geometry in solution (m-c).
This X-point geometry of the poloidal magnetic field has not been seen 
in  previous equilibrium studies
(\citealt{Glampedakis_Lander_Andersson_2014}; \citealt{Vigano_Pons_Miralles_2011}) 
nor in  the equilibrium models produced by a  numerical simulation study (\citealt{Parfrey_Beloborodov_Hui_2013}).
The X-point geometry would be unstable because it would cause 
magnetic reconnection as \cite{Parfrey_Beloborodov_Hui_2013} calculated.
Our numerical result shows that the X-point geometry appears 
when the strength of the equatorial shearing (magnetospheric toroidal current density) exceeds
 a certain value.

\begin{table*}
 \begin{tabular}{cccccccccccccccc}
\hline
&  $\hat{r}_M$ & $\hat{I}_0$ & $\hat{S}_0$ & $\hat{F}_0$ & model & $E{\rm cr}_t / E$ & $E{\rm co}_t / E$ &
 $E{\rm cr} / E$ & $E{\rm co} / E$ &   $E{\rm st}/E$ &   $ \hat{E}$ & $\hat{B}_d$ & $\hat{r}_X$ & $\hat{j}_0$  \\
\hline
(s-a) & 8.0   &  1.1 & -1  & 0.5  & IV  & 5.89E-2 & 2.50E-2 & 0.03 & 0.86 & 0.89 & 2.58E-2 & 1.70E-1 &   -  & 0.0 \\
(s-b) & 8.0   &  1.1 & -1  & 0.5  & IV  & 7.06E-2 & 1.56E-2 & 0.06 & 0.67 & 0.73 & 1.42E-1 & 6.15E-1 &   -  & 0.5 \\
(s-c) & 8.0   &  1.1 & -1  & 0.5  & IV  & 7.20E-2 & 1.36E-2 & 0.07 & 0.65 & 0.72 & 2.75E-1 & 8.82E-1 & 3.34 & 0.8 \\
\hline
(e-a) &  4.0   &   2.5  & -1.0    & 0.5   & IV   & 8.14E-2 & 3.50E-2 & 0.03 & 0.80 & 0.83  & 3.32E-2 & 2.10E-1 & 1.59  \\
(e-b) &  8.0   &   1.2  & -1.0    & 0.5   & IV   & 6.08E-2 & 2.71E-2 & 0.03 & 0.85 & 0.88  & 2.65E-2 & 1.73E-1 & 3.29  \\
(e-c) & 15.0   &   0.75 & -1.0    & 0.5   & IV   & 3.53E-2 & 1.53E-2 & 0.03 & 0.87 & 0.90  & 2.38E-2 & 1.56E-1 & 6.04  \\
  \hline
 \end{tabular}
\caption{Parameters and solutions with a equatorial models. 
Solutions with current sheet -- (s-a, s-b, s-c) -- and
without a current sheet changing the value of $\hat{r}_M$ -- (e-a), (e-b), (e-c) -- are tabulated.}
 \label{Tab:tab2}
\end{table*}

\begin{figure*}
 \begin{center}
  \includegraphics[width=6cm]{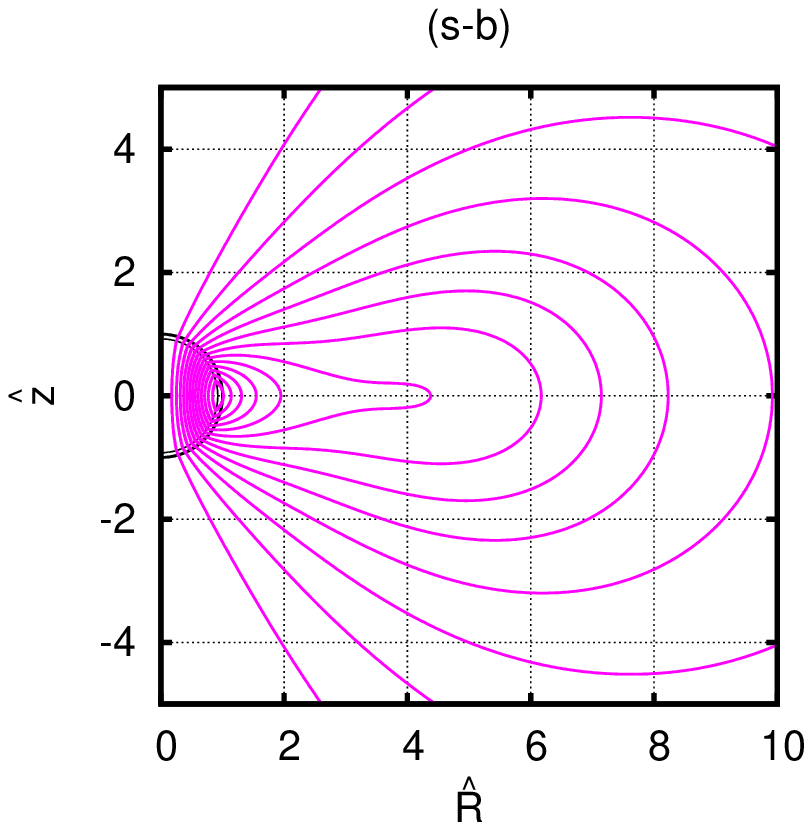}
  \includegraphics[width=6cm]{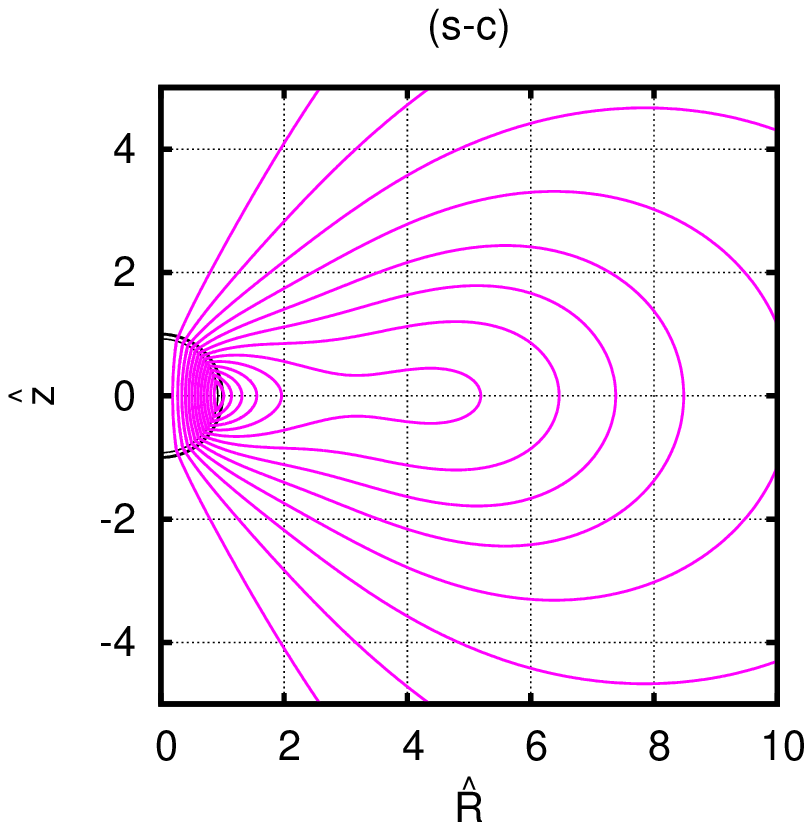}
 \end{center}

\caption{The contour of $\hat{\Psi}$. 
 The inner circles in the top panels denote stellar surfaces. An 
X-point geometry appears at $\hat{r} \sim 3.34$  in solution (s-c).}
\label{Fig:fig4}
\end{figure*}

 \begin{figure*}
  \begin{center}
   \includegraphics[width=8cm]{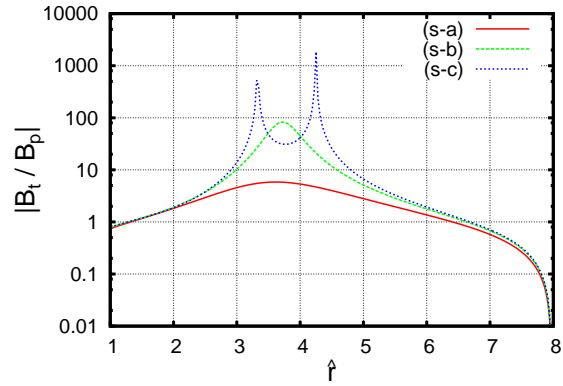}
   \caption{The ratios of the toroidal component to the poloidal component $|B_t/B_p|$.
   The figure shows solution (s-a) (solid line), 
   solution (s-b) (dashed line), and solution (s-c) (dotted line).}
   \label{Fig:BtBp2}
  \end{center}
 \end{figure*}

 \begin{figure*}
  \includegraphics[width=8cm]{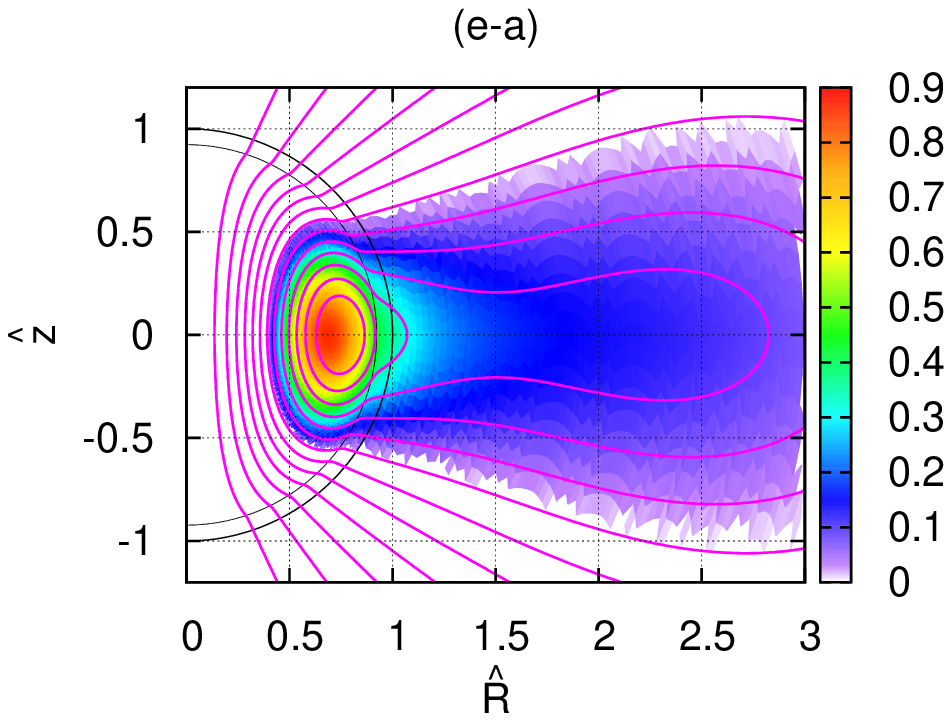}
  \includegraphics[width=8cm]{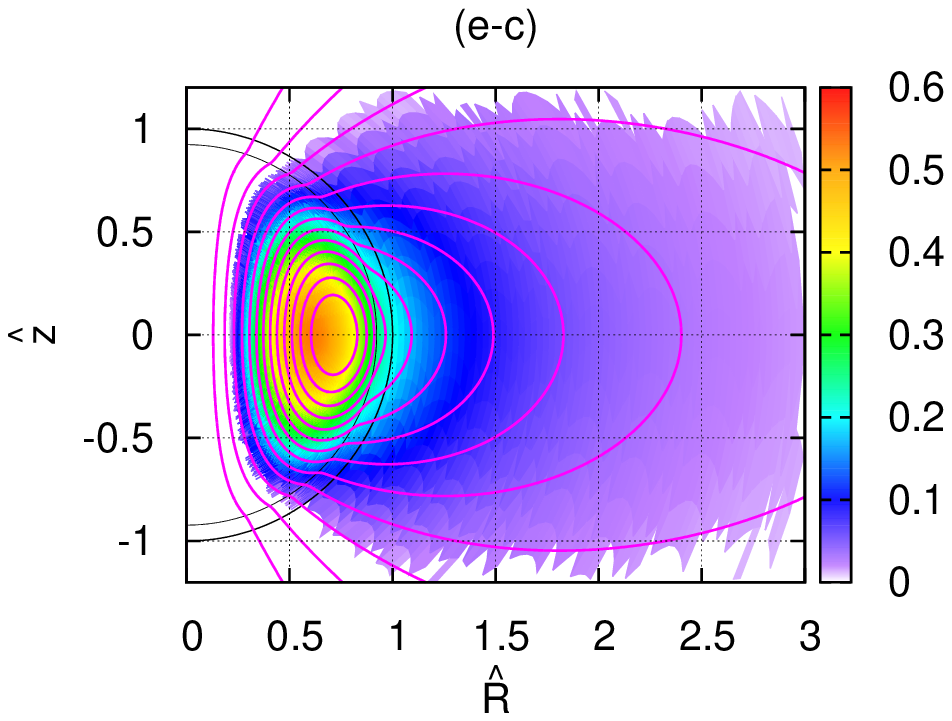}

\caption{The counter of $\hat{\Psi}$ (solid line) and 
the distribution of $B_\varphi$ (colour map) normalized by the 
magnetic dipole component strength at the north pole.
The toroidal magnetic field region within the crust becomes larger
as the size of the magnetosphere increases.
Both solution (e-a) and solution (e-c) have X-point geometry  
fields near $\hat{r} \sim 1.59$ for solution (e-a) and $\hat{r} \sim 6.04$ for solution (e-c).}
\label{Fig:fig2}
  \end{figure*}

Next, we calculated solutions with a current sheet on the core-crust interface
and examined the influence of the current sheet on the X-point geometry.
The X-point geometry is formed by the strong magnetospheric twisted field
(magnetospheric toroidal current density) compared to 
the stellar magnetic fields (stellar toroidal current density)
(see the values $E{\rm st}/E$ in Tab. \ref{Tab:tab2}).  
As we have seen, 
the large twisted field causes the X-point geometry as in solutions (m-c) and (m-f).
On the other hand, the current sheet on the core-crust interface 
changes the stellar magnetic fields as shown in Sec. 3.1.
Such current sheets can produce the X-point geometry.
We calculated the solutions for both positive and negative current sheets
and found that a positive current sheet tends to form an X-point geometry.
We show three solutions -- (s-a), (s-b), ans (s-c) with model IV core magnetic fields.
We calculated them using the same parameters, apart from  the
strength of the current sheet on the core-crust boundary $\hat{j}_0$.
The numerical results are tabulated in 
the upper column of Tab. \ref{Tab:tab2} and are displayed in Fig. \ref{Fig:fig4}. 
Fig. \ref{Fig:BtBp2} shows the ratio of $B_t$ to $B_p$ on the equatorial surface. 
The maximum value of $|B_t / B_p|$ for  solution (s-b) is approximately $100$.
This value is much larger than that of solution (m-e) which 
has the same internal field model and no current sheet.
From Fig. \ref{Fig:BtBp}, the critical value of  $|B_t / B_p|$ would be approximately $100$.

As seen in Tab. \ref{Tab:tab2}, the positive
current sheet decreases the stellar magnetic field energy ratio (see $Est / E$).
Thus, the value of $E - Est$ is increased by the positive current sheet.
As a result, an X-point geometry is formed within the twisted region 
in solutions (s-c) (see Fig. \ref{Fig:BtBp2}).
The positive current sheet has magnetic pressure as solution (d) in Fig. \ref{Fig:fig1}.
These numerical results imply that the core-crust boundary condition and the magnetic pressure
can change the magnetospheric configurations and
cause an X-point geometry within the twisted region.  
We discuss X-point geometry in Sec. \ref{subsec:magnetar_flare}.

Finally, we observed the internal magnetic field configurations
changing the size of the twisted-magnetosphere (the value of $\hat{r}_M$).
We calculated three solutions 
(e-a) ($\hat{r}_M = 4.0$), (e-b) ($\hat{r}_M = 8.0$), and (e-c) ($\hat{r}_M = 15.0$)
with model IV internal magnetic fields.
The numerical solutions are tabulated in the lower column of Tab. \ref{Tab:tab2}
and are displayed in Fig. \ref{Fig:fig2}.
Figure \ref{Fig:fig2} shows the contours of $\hat{\Psi}$ (poloidal magnetic field) 
and $B_\varphi$ (colour map) normalized by the strength of the surface dipole field.
As seen in Fig. \ref{Fig:fig2}, 
the size of the toroidal magnetic field region within the crust 
becomes much larger than those without a magnetosphere (compare with Fig. \ref{Fig:fig1}). 
As the size of the twisted field increases ( sequentially  from (e-a) to (e-c) ), 
the size of the toroidal magnetic field region within the crust also increases. 
As seen in Tab. \ref{Tab:tab2}, however, 
the energy ratios $E{\rm{cr}_t}/ E{\rm{cr}}$ and $E{\rm co}_t / E{\rm co}$ decrease
as $\hat{r}_M$ increases. 
Since the boundary condition of $\hat{I}$ on the stellar surface is 
$\hat{I} \neq 0$ in the region ($\hat{\Psi} > \hat{\Psi}_{t,\max}$),
the boundary condition $\hat{I} \neq 0$ significantly  broadens the 
size of the toroidal magnetic field region within the crust.

All solutions (e-a), (e-b), and (e-c) have X-point geometries.
The energy ratio $Est/E$ increases from 0.83 (e-a) to 0.90 (e-c) 
as the size of the twisted field region becomes large.
In other words,  the value of $E - E{\rm st}$ for solution (e-a) is 
larger  than for solutions (e-b) and (e-c).
This result implies that the star requires a stronger 
magnetospheric energy (twist) to make an
X-point geometry near the stellar surface as in solution (e-a).
We see this tendency in following subsection.

\subsection{Solutions with a ring shearing magnetosphere}

\begin{table*}
 \begin{tabular}{cccccccccccccccc}
\hline
&  $\hat{r}_M$ & $\hat{I}_0$ & $\hat{I}_1$ & $\hat{S}_0$ & $\hat{F}_0$ & model 
&$E{\rm ct}_t / E{\rm cr}$ & $E{\rm co}_t / E{\rm co}$ 
&  $E{\rm cr}/ E$ & $E{\rm co} / E$ & $E{\rm st}/E$   &  $\hat{E}$ & $\hat{B}_d$ & $\hat{r}_X $ \\
\hline
(r-a)&   1.0   & 10 & 0.0  &-1.0     & 0.1   & IV & 7.37E-3 & 2.68E-1 & 0.02 & 0.95 & 0.97 & 9.85E-3 & 4.75E-2 & - \\
(r-b)&   4.0   & 10 & 250  &-1.0     & 0.1   & IV & 7.82E-3 & 2.68E-1 & 0.02 & 0.95 & 0.97 & 9.76E-3 & 4.74E-2 & - \\
(r-c)&   8.0   & 10 &120  &-1.0     & 0.1   & IV   & 7.53E-3 & 2.68E-1 & 0.02 & 0.95 & 0.97 & 9.76E-3 & 4.74E-2 & -  \\
(r-d)&  15.0   & 10 & 80.0 &-1.0     & 0.1   & IV   & 7.48E-3 & 2.68E-1 & 0.02 & 0.95 & 0.97 & 9.80E-3 & 4.74E-2 & 6.33 \\
\hline
 \end{tabular}
\caption{Parameters and solutions of ring shearing models
(r-a), (r-b), (r-c),  and (r-d) .}
 \label{Tab:tab3}
\end{table*}

\begin{figure*}
 \begin{center}
 \includegraphics[width=8cm]{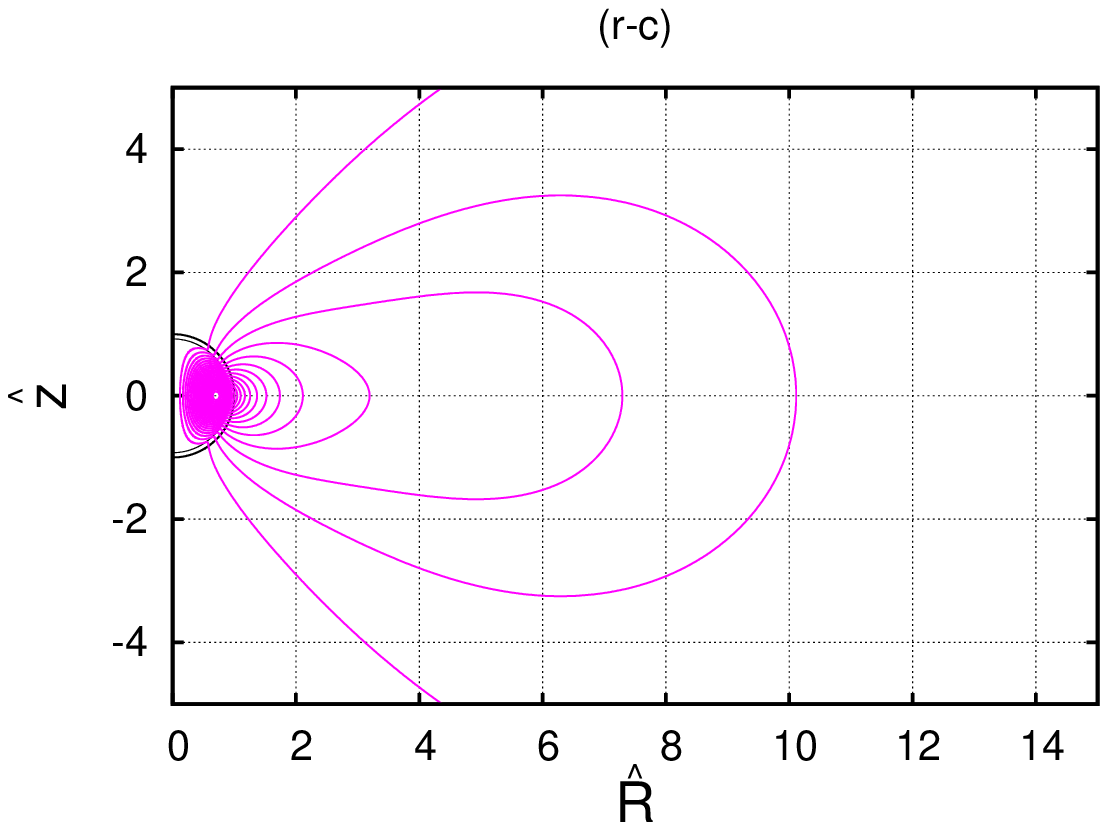}
 \includegraphics[width=8cm]{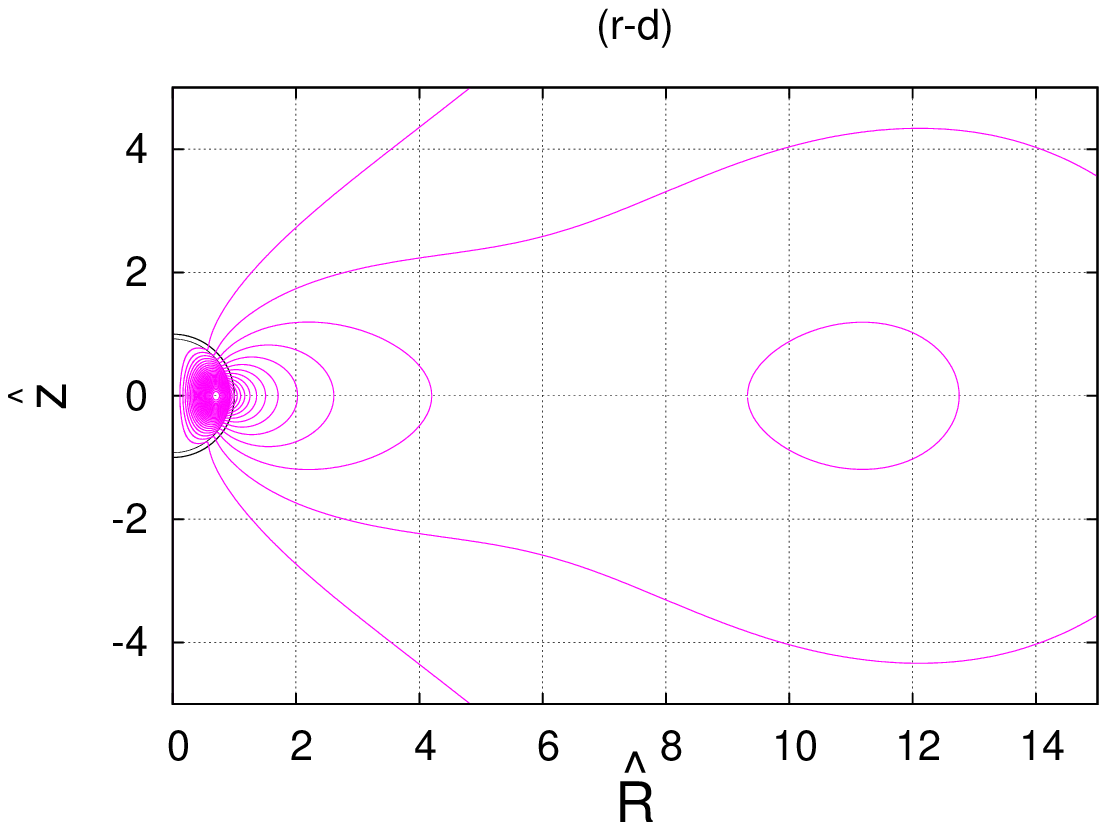}
 
 \includegraphics[width=8cm]{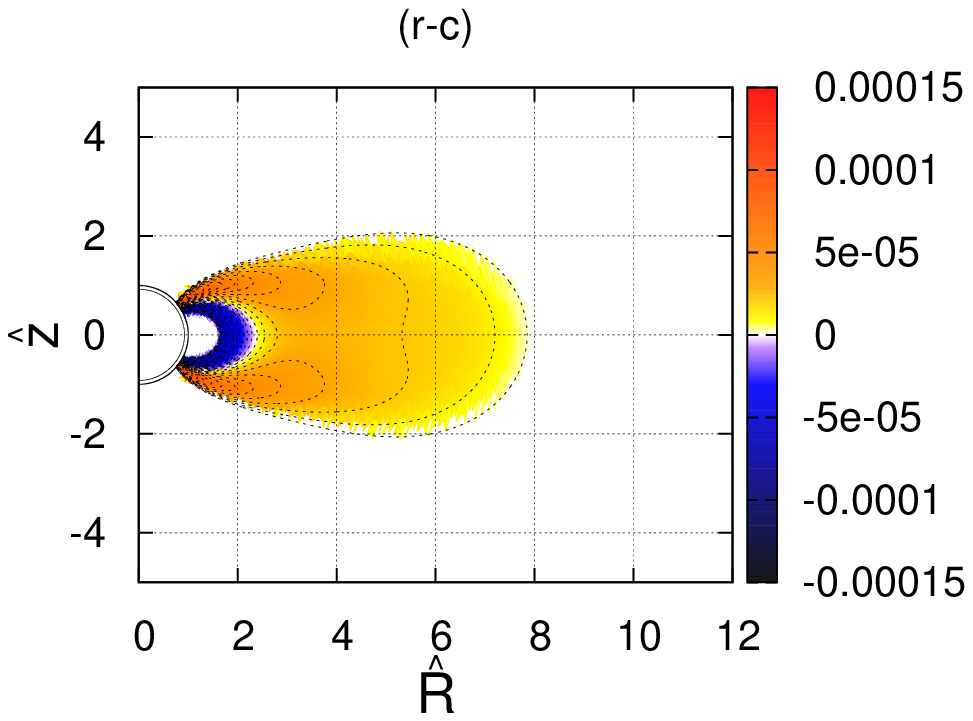}
 \includegraphics[width=8cm]{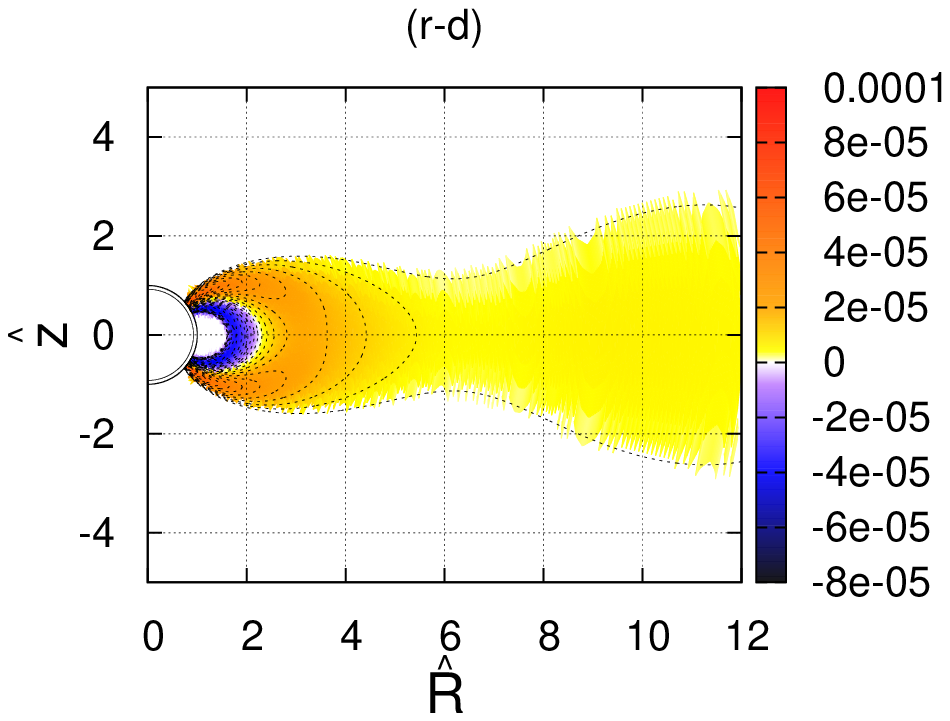}
 \end{center}
\caption{Top panels: The contour of $\hat{\Psi}$.
Bottom panels: The colour maps and contours of $\hat{j}$ in  a twisted magnetosphere.
 The inner circles in all panels denote stellar surfaces.  
While  an X-point geometry appears $\hat{r} \sim 6.33$ in solution (r-d) (right panels),
solution (r-c) do not have X-point geometry (left panel).
The ring sharing models have both positive and negative toroidal current density in the magnetosphere.
}
 \label{Fig:fig3}
\end{figure*}

Finally, we  calculated magnetized equilibria of model IV internal magnetic 
fields with a ring shearing magnetosphere (\citealt{Parfrey_Beloborodov_Hui_2013}). 
The numerical results are tabulated in Tab. \ref{Tab:tab3} 
and are shown in Fig. \ref{Fig:fig3}. 
The twisted region is limited between the field lines $\hat{\Psi} = \hat{\Psi}_{t,\max}$ 
and $\hat{\Psi} = \hat{\epsilon} \hat{\Psi}_{ex,\max}$ (see Eq. \ref{Eq:I2}).
The toroidal current density of the ring model has both positive and negative values 
(see bottom panels in Fig. \ref{Fig:fig3}) in the magnetosphere. 
This distribution is similar to the solutions by \cite{Parfrey_Beloborodov_Hui_2013} 
(see figure 4 in their paper). 
The energy ratio $E{\rm cr}_t / E{\rm cr} $ of solutions (r-b), (r-c), and (r-d) 
are almost the same as for solution (r-a).
The magnetospheric toroidal current density near the star
is much smaller than that of the equatorial model (see solution (e-c) in Fig. \ref{Fig:fig2})
because the magnetospheric toroidal current is limited within the narrow regions
and takes both positive and negative values.
The ring shearing magnetosphere in this paper
does not significantly influence the internal 
magnetic field configurations.

We cannot obtain solutions with X-point geometry 
within $\hat{r}_M < 15$  in this paper
because the magnetospheric toroidal current 
density within the ring shearing model is 
much smaller than in the equatorial shearing model.
Solutions (r-b) and (r-c) do not have X-point geometries.
A large size of the twisted region is required in order to
form the X-point geometry.
An X-point geometry appears only when the size 
of the twisted field region is sufficiently large such as in solution (r-d) ($\hat{r}_M = 15$)
in our numerical results.
These results show that an X-point 
geometry of the ring shearing model in this paper cannot appear near 
the stellar surface, unlike for the equatorial shearing models. 
The X-point geometry in the ring shearing model tends to 
appear in the outer region of the equatorial shearing  X-point geometry.
The minimum value of $\hat{r}_X$ in the ring shearing model is larger 
than for the equatorial shearing model in these calculations. 
This result means that the minimum value of $\hat{r}_X$ 
changes according to the shearing model and 
the size of the twisted field in the magnetosphere. 

\section{Discussion and concluding remarks}

\subsection{The influence of boundary conditions on the Hall MHD evolution}
\label{subsec:MHD_evolution}

The boundary conditions of the crust play very important 
roles in the Hall MHD secular evolutions.
Very recently, \cite{Vigano_et_al_2013} performed Hall MHD simulations
under three types of core magnetic fields models 
(see Fig.2 in \citealt{Vigano_et_al_2013}).
They showed that the evolutionary path 
is affected by the core magnetic field. 
We will now discuss the influence of the core magnetic fields 
on the crustal magnetic fields in Hall equilibrium  
using our equilibrium solutions.

We calculated 6 solutions 
and found that the Hall equilibrium state within the crust is significantly affected
 by the core-crust and the stellar surface-magnetosphere boundary conditions.
As we have seen in Sec. \ref{Sec:core_magnetic}, 
the presence of the core magnetic field changes 
the crustal magnetic field in the Hall equilibrium (solutions (c)).
Moreover, the strong core toroidal magnetic field 
broadens the regions of the crustal toroidal magnetic field in the Hall equilibrium (solution (e)). 
The surface boundary condition also changes the crustal magnetic field 
configurations. Additionally,  the twisted magnetosphere around the star increases 
the size of the crustal toroidal magnetic fields.
These boundary condition 
affect the Hall cascade during the secular magnetic field evolution,
because the dissipation timescale of the current sheet 
is larger than the Hall timescale (App. \ref{App:dissipation}).

The initial large toroidal magnetic fields strengthen 
the Hall drift activity (\citealt{Kojima_Kisaka_2012}; \citealt{Gourgouliatos_Cumming_2014}).
If the crustal magnetic fields are large enough to drive the Hall cascade very effectively ($\geq 10^{15}$G), 
the timescale of the Hall drift is much faster than the Ohmic timescale (see Eq. \ref{Eq:R_m}).
In this case, the Hall drift would stop when the 
initial toroidal magnetic fields have been changed into
poloidal components by the Hall cascade and have 
reached Hall equilibrium configurations. 
Therefore, the energy ratio $E{\rm cr}_t / E{\rm cr}$ of the Hall equilibria would 
represent the efficiency of the Hall cascade
if the magnetar has strong initial toroidal component magnetic fields 
at the beginning of the Hall MHD secular evolution.

The initial magnetic field configurations in the magnetar interior
are still unclear, but they would have strong 
toroidal magnetic fields because of  the rapid differential rotation 
(\citealt{Duncan_Thompson_1992}; \citealt{Spruit_IAU259}).
The magnetar would be in  a stable MHD equilibrium state 
 because the Alfv\'en timescale ($= \sqrt{4\pi 
\rho}r_s / B \sim 0.1$s for a typical magnetar 
with magnetic fields in the order of $B=10^{15}$G)
is much faster than the dynamical and crust forming timescales.
The MHD equilibria with strong toroidal magnetic 
fields would be  dynamically stable because they  satisfy
Braithwaite's stability criterion
(\citealt{Braithwaite_2009}; \citealt{Duez_Braithwaite_Mathis_2010}):
\begin{eqnarray}
 a\frac{E}{|W|} < \frac{E_p}{E} \leq 0.8, 
\label{Eq:criteria}
\end{eqnarray}
where $E/|W|$ is the ratio of magnetic energy to 
gravitational energy. $a$ is a certain dimensionless factor 
in the order of $10^3$ for neutron stars. Since the value of $E/|W|$ 
is smaller than $\sim 10^{-5}$ for typical magnetars,
the left-hand side of this inequality (Eq. \ref{Eq:criteria}) could be less than approximately
$10^{-2}$. Therefore, this criterion means 
the configurations 
are
dynamically stable even if the toroidal magnetic fields are much stronger than
the poloidal magnetic fields.
As a result, it is natural that the magnetar
has strong initial toroidal magnetic fields in its interior.

Here, we assume that the magnetar has very large strong initial 
toroidal magnetic fields and that the toroidal magnetic fields
are changed into poloidal components by the Hall cascade
until the system reaches the Hall equilibrium state.
This situation is different from initial models 
in \cite{Vigano_et_al_2013} because except for A14T in their paper, 
their initial models  do not have toroidal magnetic fields.
If the value of $E{\rm cr}_t / E{\rm cr}$ in the Hall equilibrium is very small ($\sim 0.1\%$),
the Hall drift would be effective because 
almost of the  initial toroidal components  
are changed into poloidal components by the Hall drift.
On the other hand, if the value of $E{\rm cr}_t / E{\rm cr}$ 
in Hall equilibrium is not small ($\sim 10\%$),
the Hall drift would be less effective. 
Therefore, we can evaluate the efficiency of the Hall drift 
using the toroidal magnetic field energy ratio of each solutions.

The Hall drift of model I, such as in solution (a),
would be very active because most toroidal magnetic field components
would change to poloidal magnetic field components.
This is consistent with the calculations by \cite{Kojima_Kisaka_2012} 
(see the left panel of Fig.5 in \citealt{Kojima_Kisaka_2012}).
In contrast, the value of the energy ratio $Ecr_t/Ecr$ of solution (c)
is much larger than that of solution (a). 
These numerical results mean that the Hall 
cascade in such configurations would be more passive 
than the configuration in solution (a).
Therefore, the presence of the core magnetic fields 
tends to weaken the Hall drift within the crust region.

The crustal magnetic field in the Hall equilibrium with 
a  twisted-magnetosphere also has a  very large 
toroidal magnetic field region. Since the size of the 
toroidal magnetic field region of the equatorial twisted model
is much larger than that of the non-twisted model, 
the Hall cascade with a twisted magnetosphere would decrease. 
We see the same tendency in Sec. 2.1.2 of \cite{Pons_Geppert_2007}.
They calculated force-free boundary models and argued 
that the Hall drift within the crust is reduced by the  twisted magnetosphere.
 
\subsection{Magnetic X-point geometry and flare}
\label{subsec:magnetar_flare}

We have calculated core-crust magnetic field configurations 
with a twisted force-free magnetosphere. We found 
interesting poloidal magnetic field configurations in the magnetosphere. 
When the magnetospheric toroidal current density
is relatively large, a magnetic X-point geometry is
formed in the equatorial plane of the magnetosphere. 
This  X-point geometry  is  constituted by anti-parallel 
poloidal magnetic fields around the point.
These anti-parallel magnetic fields would cause a magnetic reconnection and 
result in a giant flare of the magnetar (
\citealt{Masada_Nagataki_Shibata_Terasawa_2010}; \citealt{Parfrey_Beloborodov_Hui_2013}).
 The distance of the X-point geometry changes as a result of  the shearing models.
 The equatorial sharing model can cause an  X-point geometry near the stellar surface
 and the minimum value of $\hat{r}_X$ is $\sim 1.59$ as in solution (e-a) (Tab.\ref{Tab:tab1}).
 In contrast, the ring sharing model cannot make an X-point geometry 
 near the stellar surface and  the minimum value of $\hat{r}_X$ is $\sim 6.33$ 
 in solution (r-d) (Tab. \ref{Tab:tab2}).
 The location of the magnetic reconnection in the magnetosphere 
 would deeply depend on the magnetospheric models. 

Next, we considered the physical process for an X-point geometry.
We  obtained an X-point geometry by changing three parameters. 
An X-point geometry appears by condition (1)  increasing the value 
of toroidal magnetic fields ($\hat{I}_0$) (solutions (m-c) and (m-f)),
condition (2)  increasing the strength of the current sheet ($\hat{j}_0$) on the
core-crust boundary (solution (s-c)), and
condition (3)  increasing the size of the twisted magnetosphere  $(\hat{r}_M)$
(the size of the surface toroidal magnetic field) (solution (r-d)).

 Condition (1) is satisfied in the case where the magnetospheric 
toroidal current is accumulated by the fixed stellar current (\citealt{Parfrey_Beloborodov_Hui_2013}). 
Then, the value of $\hat{I}_0$ increases as the energy 
of the toroidal magnetic field in the magnetosphere increases.
Condition (2) is satisfied in the case where 
the internal magnetic field configurations and elastic force balance in the crust 
are changed by dynamical events such as a glitch or elastic crust cracking.
Since these events would change the core-crust boundary condition,
the value of $\hat{j}_0$ varies in this case as 
we have changed the current sheet on the boundary.
Condition (3) is satisfied in the case where
the toroidal magnetic fields emerge from the crust.
The size of the surface toroidal magnetic fields 
(twisted magnetosphere) is enlarged because we have changed the value of $\hat{r}_M$.
The large crust cracking would result in the toroidal magnetic fields 
emerging.  Such cracking releases stress on the crust 
and decreases the negative current sheet on the core-crust boundary.
During this process,  the internal magnetic field configurations would change from
a configuration as per solution (f) to a configuration as per solution (d).
The direction of the magnetic pressure also changes as seen in Fig. \ref{Fig:fig1}.
The strength of the negative current sheet ($\hat{j}_0$) decreases 
and the the strength of the positive current sheet increases during this process.
Since the energy ratio $E{\rm co}_t/E{\rm co}$ decreases from solutions (f) to (d), 
the core toroidal magnetic field would be ejected 
from the star (\citealt{Thompson_Duncan_2001}; \citealt{Fujisawa_Eriguchi_2013}). 
At the same time, the twist of the magnetosphere would increase 
as seen in Fig. \ref{Fig:fig4}, because the strength of the 
positive current sheet ($\hat{j}_0$) is increasing. It is probable that
such toroidal magnetic energy injection would make
giant magnetic loops (\citealt{Gourgouliatos_Lynden-Bell_2008}; 
\citealt{Takahashi_Asano_Matsumoto_2009, Takahashi_Asano_Matsumoto_2011};
\citealt{Matsumoto_Masada_Asano_Shibata_2011}).
These dynamical events in the stellar interior and the changes of the toroidal magnetic field
 on the surface would result in a reconnection within the magnetosphere and 
a giant flare. Our numerical results would be a key to understanding
the physical mechanisms of a magnetar's giant flare.

\subsection{Concluding remarks}
 
We systematically and simultaneously  calculated magnetized equilibria
throughout an MHD equilibrium core, Hall equilibrium crust,
and twisted force-free magnetosphere
with both poloidal and toroidal magnetic 
fields using the Green function method
under various boundary conditions.
We employed SLy EOS as a realistic equation of state
in order to treat the core and crust consistently.
We found that the magnetic field
configuration in each region is 
deeply affected by the other.

We found that the Hall equilibrium state is significantly affected by 
both inner (core) and outer (twisted-magnetosphere) boundary 
conditions. The core magnetic field 
configurations influence  the 
crustal toroidal magnetic field.
The crustal magnetic fields, such as in 
 the model with a purely crustal magnetic field (solution (a))
have the smallest toroidal magnetic field energy ratio
among all of our solutions. 
In contrast, 
the core magnetic fields of solutions 
such as solution (c) sustain the larger 
toroidal magnetic field energy ratio ($E{\rm cr}_t / E{\rm cr}$)
within the crust in the Hall equilibrium. 
The twisted force-free magnetosphere 
also widens the size of the crustal toroidal magnetic field 
region. We can evaluate the efficiency of the Hall cascade 
using the toroidal magnetic field energy ratio $E{\rm cr}_t / E{\rm cr}$
 in the Hall equilibrium.  The presence of the core magnetic fields would 
weaken the efficiency of the Hall cascade
because the crustal toroidal magnetic field energy is not small
when the crustal magnetic fields reach the Hall equilibrium.
Since the twisted magnetosphere widens the
size of the crustal toroidal magnetic field region, 
it also weakens the Hall cascade within the crust.

The magnetosphere around the star forms a magnetic X-point geometry 
when the magnetospheric toroidal current density is 
sufficiently large or the stellar total current is sufficiently small. 
A magnetic X-point geometry can be caused by a 
physical event between the core and crust 
such as a glitch or a changing magnetic field. 
An X-point geometry causes a magnetic reconnection 
which would be an origin of a giant flare of a magnetar.
The location of an X-point geometry depends on the 
size of the twisted region and the shearing model. Equatorial shearing 
causes an X-point geometry near the stellar surface ($\hat{r}_X \sim 1.59$), 
but the ring sharing cannot make the 
X-point geometry near the surface. The X-point geometry in the 
ring model appears at a  distant region near $\hat{r}_X \sim 6.33$. 
The critical ratio of the poloidal component to toroidal component is estimated
to be approximately $|B_t /B_p| \sim 100$.
These numerical results show that 
both Hall MHD secular evolution and 
magnetospheric dynamical evolution 
would be affected by the magnetic field configurations of
other regions and the core-crust boundary conditions.
We need to consider core, crustal, and magnetospheric 
magnetic field configurations simultaneously 
in order to investigate magnetar physics. 

\section*{ACKNOWLEDGEMENTS}

The authors would like to thank an anonymous reviewer for useful
comments and suggestions that helped us to improve this paper.
KF would like to thank the members of plasma seminar at the NAOJ
for the very exciting discussion and drinks after the seminar.
KF would like to thank Prof. T. Terasawa 
and Prof. S. Yamada for their valuable and useful comments.
KF was supported by Grant-in-aid for JSPS Fellows.
KF and SK are supported by Grant-in-Aid for 
Scientific Research on Innovative Areas, No. 24103006.

\bibliographystyle{mn}

\end{multicols}


\appendix

\section{Derivations of the basic equations}
\label{App:eqs}

The evolution of the magnetic field is governed by the induction equation
 \begin{eqnarray}
  \P{}{t} \Vec{B} = - c\nabla \times \Vec{E}.
 \end{eqnarray}
 The electrical field in the crust is expressed as (\citealt{Goldreich_Reisenegger_1992}),
 \begin{eqnarray}
  \Vec{E} = \frac{\Vec{j}}{\sigma} + 
 \frac{1}{e n_e c} \Big(\Vec{B} \times \Vec{j} \Big) + \nabla \mu,
 \end{eqnarray}
 where $\sigma$ is the electric conductivity, $n_e$ is the electron number density, 
 $e$ is the charge of an electron and $c$ is the speed of light.
 $\nabla \mu$ denotes the gradient of the total chemical potential term.
 Therefore, the Hall MHD evolutionary equation is
 \begin{eqnarray}
  \P{}{t} \Vec{B} = - \nabla \times \left(\frac{c^2}{4\pi \sigma} \nabla \times \Vec{B}  \right)
  + \nabla \times \Big[\frac{c}{4\pi e n_e} \Vec{B} \times (\nabla \times \Vec{B}) \Big].
 \label{Eq:Hall_MHD}
 \end{eqnarray}
 The first term on the right hand side  is the Ohmic diffusion and the second is the Hall drift.
 If the Hall drift is much faster than the Ohmic diffusion and the magnetic field reaches 
 nearly stationary, the left hand side and the first term on the right hand side of the equation vanishes.
 The evolutionary equation becomes the Hall equilibrium condition as follows: 
\begin{eqnarray}
 \nabla \times \Big[\frac{c}{4\pi e n_e} \Vec{B} \times (\nabla \times \Vec{B}) \Big] = 0.
\end{eqnarray}
Since $c$ and $e$ are physical constants,
this implies the presence of a scalar function $S$ as
\begin{eqnarray}
 \nabla S = \frac{1}{n_e} \left( \Vec{B} \times \frac{\Vec{j}}{c} \right).
 \label{Eq:conditions}
\end{eqnarray}
From the axisymmetry condition, we obtain
\begin{eqnarray}
 \frac{1}{n_e} \left( \Vec{B} \times \frac{\Vec{j}}{c} \right)_\varphi = 0.
  \label{Eq:condition_varphi}
\end{eqnarray}
These conditions are the same as the stationary conditions in the 
barotropic axisymmetric MHD except for $n_e$. 
The physical dimensions of these equations are different, but
the physical meaning is almost identical. 
The Hall drift term originates from the Lorentz force 
of the electron.
Therefore, we obtain  the following condition
from the toroidal component of Eq. (\ref{Eq:condition_varphi})
\begin{eqnarray}
 \nabla \Psi \times \nabla I = 0 \Leftrightarrow I \equiv I(\Psi).
  \label{Eq:condition_pol}
\end{eqnarray}
From the meridional component of Eq. (\ref{Eq:conditions}), the functional 
form of the toroidal current density
is obtained as
\begin{eqnarray}
4 \pi \frac{j_\varphi}{c} = \frac{I(\Psi) I'(\Psi)}{r\sin \theta} + 4 \pi n_e r \sin \theta S(\Psi),
\end{eqnarray}
where $I' = \D{I}{\Psi}$. 

\section{Generating function for the Legendre polynomial}
\label{App:Legendre}

The solution of the Poisson equation,
\begin{eqnarray}
\Delta ( A(r,\theta) \sin \varphi) = -4\pi j(r,\theta) \sin \varphi,
\label{App:Poisson_eq}
\end{eqnarray}
is expressed by the $m=1$ Legendre function $P_n^1 (\cos \theta)$ 
in an axisymmetric system as follows:
{\small
\begin{eqnarray} 
&& A(r, \theta) \sin \varphi = \sum_{n=1}^{\infty} \frac{2P_n^1(\cos \theta)}{n(n+1)} 
\int_0^\infty r'^2 f_n(r,r') dr' \int_0^\pi \sin \theta P_n^1 (\cos \theta') d\theta' 
\int_0^{2\pi} \cos (\varphi - \varphi') \, d \varphi'
j(r', \theta') \sin \varphi' \nonumber \\
&+& \sum_{n=1}^{\infty} \Big(a_n r^{n+1} P_n^1 (\cos \theta) + b_n r^{-n} P_n^1 (\cos \theta) \Big) \sin \varphi,
\nonumber \\  
&\Rightarrow& 
 A(r,\theta) = 2\pi \sum_{n=1}^{\infty} \frac{P_n^1(\cos \theta)}{n(n+1)} 
\int_0^\infty r'^2 f_n(r,r') dr' \int_0^\pi \sin \theta P_n^1 (\cos \theta') d\theta' j(r', \theta'), \nonumber \\
&+& \sum_{n=1}^{\infty} \Big(a_n r^{n+1} P_n^1 (\cos \theta) + b_n r^{-n} P_n^1 (\cos \theta) \Big),
\end{eqnarray}
}
where $P_n^1$ is the $n$th associated Legendre function and 
$f_n(r,r')$ is the following function:
\renewcommand{\arraystretch}{2}
\begin{eqnarray}
 f_n(r,r') = 
\left\{
\begin{array}{cc}
 \dfrac{1}{r}\left(\dfrac{r'}{r}\right)^n  & (r \geq r')\\
 \dfrac{1}{r'}\left(\dfrac{r}{r'}\right)^n & (r'>r) \\
\end{array}
\right.,
\end{eqnarray}
\renewcommand{\arraystretch}{1}
and the last terms  of the right hand side are homogeneous general solutions to Eq.(\ref{App:Poisson_eq}).

\section{Dimensionless forms}
\label{App:dimensionless}

Using four quantities ($r_s$, $\rho_{\max}$, $n_c$ and $S_{\max}$), 
we obtain  dimensionless forms 
of physical quantities as follows:
\begin{eqnarray}
 \hat{F} = \frac{F}{n_{c} |S_{\max}| / \rho_{\max}},
\end{eqnarray}
\begin{eqnarray}
 \hat{B} = \frac{B}{r_s^2 n_{c} |S_{\max}|},
\end{eqnarray}
\begin{eqnarray}
 \hat{\Psi} = \frac{\Psi}{r_s^4 n_{c} |S_{\max}|},
\end{eqnarray}
\begin{eqnarray}
 \hat{I} = \frac{I}{r_s^5 n_{c} |S_{\max}|},
\end{eqnarray}
\begin{eqnarray}
 \hat{j} = \frac{j_\varphi / c}{r_s n_{c} |S_{\max}|}.
\label{Eq:hat_j}
\end{eqnarray}
The functional forms of $F$ and $S$ also become dimensionless as shown below:
\begin{eqnarray}
\hat{S}(\hat{\Psi}) = \hat{S}_0,
\end{eqnarray}
\begin{eqnarray}
\hat{F}(\hat{\Psi}) = \hat{F}_0.
\end{eqnarray}
In particular, noted that the value of $\hat{S}_0$ is $1$ or $-1$ 
from its definition.

\section{Analytical solutions}
\label{App:analytical}

We show here the analytical solution of each model.
When the magnetic field is purely poloidal ($\hat{I}=0$)
and the functional forms  are $\hat{S}=\hat{S}_0$ and  $\hat{F} = \hat{F}_0$,
we can easily calculate analytical solutions. We can 
obtain the exact solutions by integrating $\hat{j}_\varphi$ throughout 
the entire region.
 We set $\hat{\rho} = \hat{\rho}_0$ and $\hat{n_e} = \hat{n}_0$,
but we can also obtain the solutions with arbitrary $\rho$ and $n_e$ distributions.
The functional form of the toroidal current density becomes
\begin{eqnarray}
 \hat{j}_\varphi = 
\left\{
\begin{array}{ll}
\hat{\rho}_0 \hat{F}_0 \hat{r} \sin \theta & (0 \leq \hat{r} < \hat{r}_{in})  \\
\hat{n}_0    \hat{S}_0 \hat{r} \sin \theta & (\hat{r}_{in} \leq \hat{r} \leq \hat{r}_{s})  \\
0                        & (\hat{r}_{s} < \hat{r}) 
\end{array}
\right.
.
 \end{eqnarray}
Now, we can obtain the analytical solutions by calculating the integral of  Eq. (\ref{Eq:Green_function}).

   \subsection{Model I}

The magnetic field within the core ($0 \leq \hat{r} < \hat{r}_{in}$) is excluded by the Meissner effect
in this model. The current density also does not exist in the core. Therefore, we fixed $\hat{F}_0 = 0$.
Then the integration of the Green function is expressed (see App. \ref{App:Legendre}) 
by 
{\small
\begin{eqnarray}
\hat{\Psi}(\hat{r},\theta) &=& 2\pi \hat{r} \sin \theta \hat{S}_0  
\sum_{n=1}^{\infty} \frac{P_n^1 (\cos \theta)}{n (n+1)} \int_0^{\infty} f_n(\hat{r},\hat{r}') \hat{r}'^2 d\hat{r}' 
 \int_0^\pi P_n^1(\cos \theta') \sin \theta' d \theta' \left(\hat{n}_0 \hat{r}'\sin \theta' \right) \nonumber \\
&+& \sum_{n=1}^{\infty} \Big(a_n \hat{r}^{n+1} P_n^1 (\cos \theta) + b_n \hat{r}^{-n} P_n^1 (\cos \theta) \Big) \sin \theta.
\end{eqnarray}
}
Since the number density profile is independent of $\theta$, we can integrate the $\theta$ component.
From the orthogonality of the associated Legendre function,
\begin{eqnarray}
 \int_0^{\pi} \sin \theta' P_n^1 (\cos \theta') P_n^1(\cos \theta') d\theta' = 2 \frac{n(n+1)}{2n+1},
\end{eqnarray}
the higher terms ($n \geq 2$) must vanish. Then, the equation becomes
{\small
\begin{eqnarray}
\hat{\Psi}(\hat{r},\theta)= \frac{4\pi}{3} \hat{r} \sin^2 \theta \hat{S}_0 \hat{n}_0 \int_0^{\infty} f_1(\hat{r},\hat{r}') \hat{r}'^3 d\hat{r}'
 + \Big(a_1 \hat{r}^{2} + b_1 \hat{r}^{-1} \Big) \sin^2 \theta.
\end{eqnarray}
}
The electrons exit within a finite region ($\hat{r}_{in} \leq \hat{r} \leq \hat{r}_{s}$) of the star.
Then, the $\hat{r}$ integral is classified into three types, -- the inner vacuum solution,  
the internal solution and the outer vacuum solution:
\renewcommand{\arraystretch}{2}
{\small
\begin{eqnarray}
\int_0^\infty f_1 (\hat{r}, \hat{r}') \hat{r}'^2 d\hat{r}' 
=\left\{
\begin{array}{ll}
\displaystyle
  \hat{r} \left(  \int_{\hat{r}_{in}}^{\hat{r}_s} \hat{r} \hat{r}' \, d\hat{r}'\right) 
= \dfrac{1}{2}\hat{r}_s^2 \hat{r}^2 - \dfrac{1}{2} \hat{r}_{in}^2 \hat{r}^2 
 & (0 \leq \hat{r} \leq \hat{r}_{in}) \\
\displaystyle
 \hat{r} \left( \int_{\hat{r}_{in}}^{\hat{r}} \dfrac{\hat{r}'^4}{\hat{r}^2} \, d\hat{r}' 
	  + \int_{\hat{r}}^{\infty} \hat{r} \hat{r}'  \, d\hat{r}'\right) 
=\left( \dfrac{1}{5}\hat{r}^4 - \dfrac{1}{5}\dfrac{\hat{r}_{in}^5}{\hat{r}} \right) 
+ \left(\dfrac{1}{2} \hat{r}_s^2 \hat{r}^2 - \dfrac{1}{2}\hat{r}^4  \right) 
 & (\hat{r}_{in} \leq \hat{r} \leq \hat{r}_{s}) \\
\displaystyle
  \hat{r} \left( \int_{\hat{r}_{in}}^{\hat{r}_s} \dfrac{\hat{r}'^4}{\hat{r}^2} \, d\hat{r}' \right)
= \left(\dfrac{1}{5} \dfrac{\hat{r}_s^5}{\hat{r}} - \dfrac{1}{5}\dfrac{\hat{r}_{in}^5}{\hat{r}}  \right)
 & (\hat{r}_{s} \leq \hat{r} )
\end{array}
\right.
.
\end{eqnarray}
}
\renewcommand{\arraystretch}{1}
If we neglect the homogeneous terms, there is no current sheet in the system
and the crustal magnetic fields penetrate the core.
Due to the Meissner effect in this model, the magnetic field cannot penetrate 
within the $\hat{r} \leq \hat{r}_{in}$ region, 
and so we must impose the boundary condition $\hat{\Psi} (\hat{r}_{in}, \theta) = 0$.
We need to choose the coefficients of the homogeneous terms in this case.
The terms represent the contribution from the toroidal current sheet on
 $\hat{r} = \hat{r}_{in}$ (\citealt{Fujisawa_Eriguchi_2013}). Therefore, 
we incorporate the toroidal current sheet in $r = r_{in}$
in order to satisfy the boundary condition $\hat{\Psi}(\hat{r}_{in}, \theta) = 0$ as follows:
\renewcommand{\arraystretch}{2}
\begin{eqnarray}
\hat{\Psi}(\hat{r},\theta) =
\left\{
\begin{array}{ll}
 \dfrac{4 \pi \hat{S}_0 \hat{n}_0}{3} \sin^2 \theta
\left[ \dfrac{1}{2} \hat{r}_s^2 \hat{r}^2 - \dfrac{1}{2} \hat{r}_{in}^2 \hat{r}^2 \right] 
+ \hat{j}_0 \hat{r}^{2}\sin^2 \theta = 0 & (0 \leq \hat{r} \leq \hat{r}_{in}) \\
 \dfrac{4 \pi \hat{S}_0 \hat{n}_0}{3} \sin^2 \theta
\left[  \left( \dfrac{1}{5} \hat{r}^4 - \dfrac{1}{5} \dfrac{\hat{r}_{in}^5}{\hat{r}} \right)
+ \left( \dfrac{1}{2} \hat{r}_s^2 \hat{r}^2 - \dfrac{1}{2} \hat{r}^4  \right)
  \right] + \hat{j}_0 \hat{r}_{in}^3 \hat{r}^{-1} \sin^2 \theta & (\hat{r}_{in} \leq \hat{r} \leq \hat{r}_s)  \\
 \dfrac{4 \pi \hat{S}_0 \hat{n}_0}{3} \sin^2 \theta
\left[  \left( \dfrac{1}{5} \dfrac{\hat{r}_{s}^5}{\hat{r}} - \dfrac{1}{5} \dfrac{\hat{r}_{in}^5}{\hat{r}} \right)
  \right] + \hat{j}_0 \hat{r}_{in}^3 \hat{r}^{-1}\sin^2 \theta  & (\hat{r}_s \leq \hat{r})
\end{array}
\right.
.
\end{eqnarray}
\renewcommand{\arraystretch}{1}
where $\hat{j}_0$ is the strength of the toroidal current sheet $(\hat{j}_0 = a_1 = \hat{r}_{in}^3 b_1)$.
Since the $a_1$ term is the inner solution and the $b_1$ term is the outer solution, 
$a_1$ is determined by the boundary condition $\hat{\Psi}(\hat{r}_{in}, \theta) = 0$ and
the explicit form of $\hat{j}_0$ is 
\begin{eqnarray}
 \hat{j}_0 = \frac{2\pi \hat{S}_0 \hat{n}_0}{3} (\hat{r}_{in}^2 - \hat{r}_{s}^2). \hspace{10pt} (0 < \hat{r}_{in})
\end{eqnarray}
Also, the sign of $\hat{j}_0$ is negative. 
This means that to exclude the magnetic field from the crustal toroidal current,
the Meissner effect equals the effective {\it negative current sheet} on the core-crust boundary 
(see Fig. 5 in \citealt{Bonazzola_Gourgoulhon_1996}).
\renewcommand{\arraystretch}{1}

\subsection{Model II}

Model II is the crustal current model, but it does not have a 
current sheet on the core-crust boundary.  Therefore, the solution is shown below:
\renewcommand{\arraystretch}{2}
\begin{eqnarray}
\hat{\Psi}(\hat{r},\theta) =
\left\{
\begin{array}{ll}
 \dfrac{4 \pi \hat{S}_0 \hat{n}_0}{3} \sin^2 \theta
\left[ \dfrac{1}{2}\hat{r}_s^2 \hat{r}^2 - \dfrac{1}{2} \hat{r}_{in}^2 \hat{r}^2 \right] & (0 \leq \hat{r} \leq \hat{r}_{in})\\
 \dfrac{4 \pi \hat{S}_0 \hat{n}_0}{3} \sin^2 \theta
\left[  \left( \dfrac{1}{5} \hat{r}^4 - \dfrac{1}{5} \dfrac{\hat{r}_{in}^5}{\hat{r}} \right) 
+ \left( \dfrac{1}{2} \hat{r}_s^2 \hat{r}^2 - \dfrac{1}{2} \hat{r}^4  \right)
  \right] & (\hat{r}_{in} \leq \hat{r} \leq \hat{r}_s)  \\
 \dfrac{4 \pi \hat{S}_0 \hat{n}_0}{3} \sin^2 \theta
\left[  \left( \dfrac{1}{5} \dfrac{\hat{r}_{s}^5}{\hat{r}} - \dfrac{1}{5} \dfrac{\hat{r}_{in}^5}{\hat{r}} \right)
  \right]  & (\hat{r}_s \leq \hat{r})
\end{array}
\right.
. 
\end{eqnarray}
\renewcommand{\arraystretch}{1}

\subsection{Models III \& IV}
\label{App:model_III_IV}

These models  have both a core current and a crustal current.
Therefore, the solutions are described by the sum of 
the core current magnetic field $\hat{\Psi}_{co}$ and the 
crustal current magnetic field $\hat{\Psi}_{cr}$.
The crustal current magnetic field $\hat{\Psi}_{cr}$ is equal to
 the model II solution. We can easily obtain the profile of $\hat{\Psi}_{co}$.
The solution is
\renewcommand{\arraystretch}{2}
\begin{eqnarray}
 \hat{\Psi}_{co}(\hat{r},\theta) = 
\left\{
\begin{array}{ll}
 \dfrac{4 \pi \hat{F}_0 \hat{\rho}_0}{3} \sin^2 \theta
\left[  \left( \dfrac{1}{5} \hat{r}^4 \right) + \left( \dfrac{1}{2} \hat{r}_{in}^2 \hat{r}^2 - \dfrac{1}{2} \hat{r}^4  \right)
  \right] & (0 \leq \hat{r} < \hat{r}_{in})  \\
 \dfrac{4 \pi \hat{F}_0 \hat{\rho}_0}{3} \sin^2 \theta
\left[  \left( \dfrac{1}{5} \dfrac{\hat{r}_{in}^5}{\hat{r}} \right)
  \right]  & (\hat{r}_{in} \leq \hat{r}) 
\end{array}
\right.
. 
\end{eqnarray}
\renewcommand{\arraystretch}{1}
As a result, the general solutions of models III and IV are
{\small
\renewcommand{\arraystretch}{2}
\begin{eqnarray}
 \hat{\Psi}(\hat{r},\theta) = 
\left\{
\begin{array}{ll}
 \dfrac{4 \pi }{3} \sin^2 \theta
\left[ \hat{F}_0 \hat{\rho}_0 \left\{ \left( \dfrac{1}{5} \hat{r}^4 \right) 
+ \left( \dfrac{1}{2} \hat{r}_{in}^2 \hat{r}^2 - \dfrac{1}{2} \hat{r}^4  \right) \right\}
       + \hat{S}_0 \hat{n}_0  \left\{  \dfrac{1}{2} \hat{r}_s^2 \hat{r}^2 - \dfrac{1}{2} \hat{r}_{in}^2 \hat{r}^2 \right\}
  \right] + \hat{j}_0 \hat{r}^2 \sin^2 \theta & (0 \leq \hat{r} < \hat{r}_{in}),  \\
 \dfrac{4 \pi}{3} \sin^2 \theta
\left[  \hat{F}_0 \hat{\rho}_0 \left( \dfrac{1}{5} \dfrac{\hat{r}_{in}^5}{\hat{r}} \right) + 
\hat{S}_0 \hat{n}_0 \left\{ \left(\dfrac{1}{5}\hat{r}^4 - \dfrac{1}{5}\dfrac{\hat{r}_{in}^5}{\hat{r}} \right) 
+ \left(\dfrac{1}{2}\hat{r}_s^2 \hat{r}^2 - \dfrac{1}{2}\hat{r}^4 \right)
\right\}
  \right] + \hat{j}_0 \hat{r}_{in}^3 \hat{r}^{-1} \sin^2 \theta  & (\hat{r}_{in} \leq \hat{r} \leq \hat{r}_s) \\
 \dfrac{4 \pi }{3} \sin^2 \theta
\left[\hat{F}_0 \hat{\rho}_0  \left( \dfrac{1}{5} \dfrac{\hat{r}_{in}^5}{\hat{r}} \right) + \hat{S}_0 \hat{n}_0 
\left\{\dfrac{1}{5} \dfrac{\hat{r}_s^5}{\hat{r}} - \dfrac{1}{5}\dfrac{\hat{r}_{in}^5}{\hat{r}}  \right\} \right]
+ \hat{j}_0 \hat{r}_{in}^3 \hat{r}^{-1} \sin^2 \theta   & (\hat{r}_{s} \leq \hat{r})
\end{array}
\right.
.
\end{eqnarray}
\renewcommand{\arraystretch}{1}
}
where $\hat{F}_0 \hat{S}_0 > 0$ (model III) and $\hat{F}_0 \hat{S}_0 < 0$ (model IV). 
Note that  models III and IV can have an arbitrary 
current sheet on the core-crust boundary. 
The last term represents the contribution from the dipole
current sheet (see \citealt{Fujisawa_Eriguchi_2013}). 
$\hat{j}_0$ denotes the strength of the current sheet.

\subsection{The magnetic field configurations of each model}

\begin{figure*}
\includegraphics[scale=0.75]{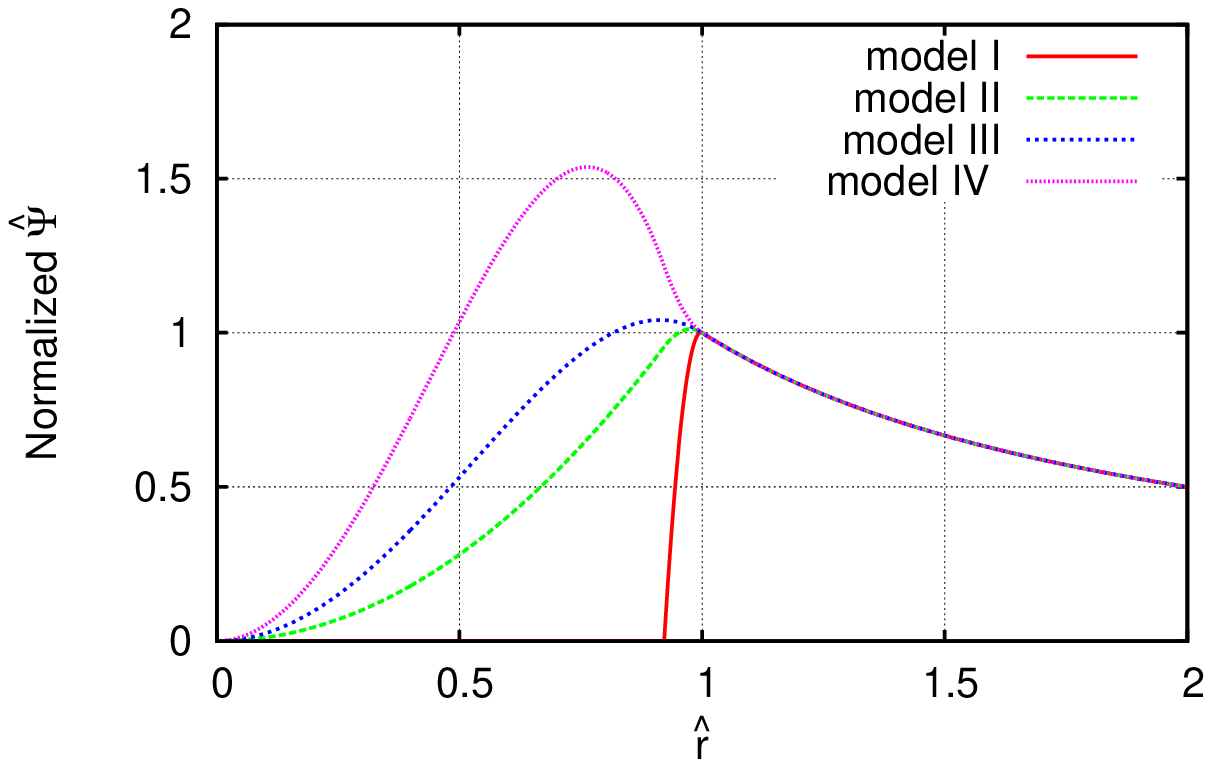} 

 \caption{The $\hat{\Psi}$ profiles of each model on the equatorial plane. 
They are normalized by the value of $\hat{\Psi}(\hat{r}_s,\pi/2)$.
 The core-crust boundary is $\hat{r} = 0.921875$.}
 \label{Fig:models_r}
 \end{figure*}

We show here some graphs of the analytical solutions.
We set $\hat{r}_s = 1$ and $\hat{r}_{in} = 0.921875$. The functional parameters are $\hat{S}_0 = 1$, $\hat{F}_0 = 1$ 
(model III) and $\hat{S}_0 = -1$, $\hat{F}_0 = 1$ (model IV). We show
 the contours of $\hat{\Psi}$ in Fig. \ref{Fig:models}.
Fig. \ref{Fig:models_r} displays profiles of $\Psi$ on the equatorial plane 
normalized by its surface value. 

\section{Accuracy verification}
\label{App:accuracy}

\begin{figure*}
 \includegraphics[width=8cm]{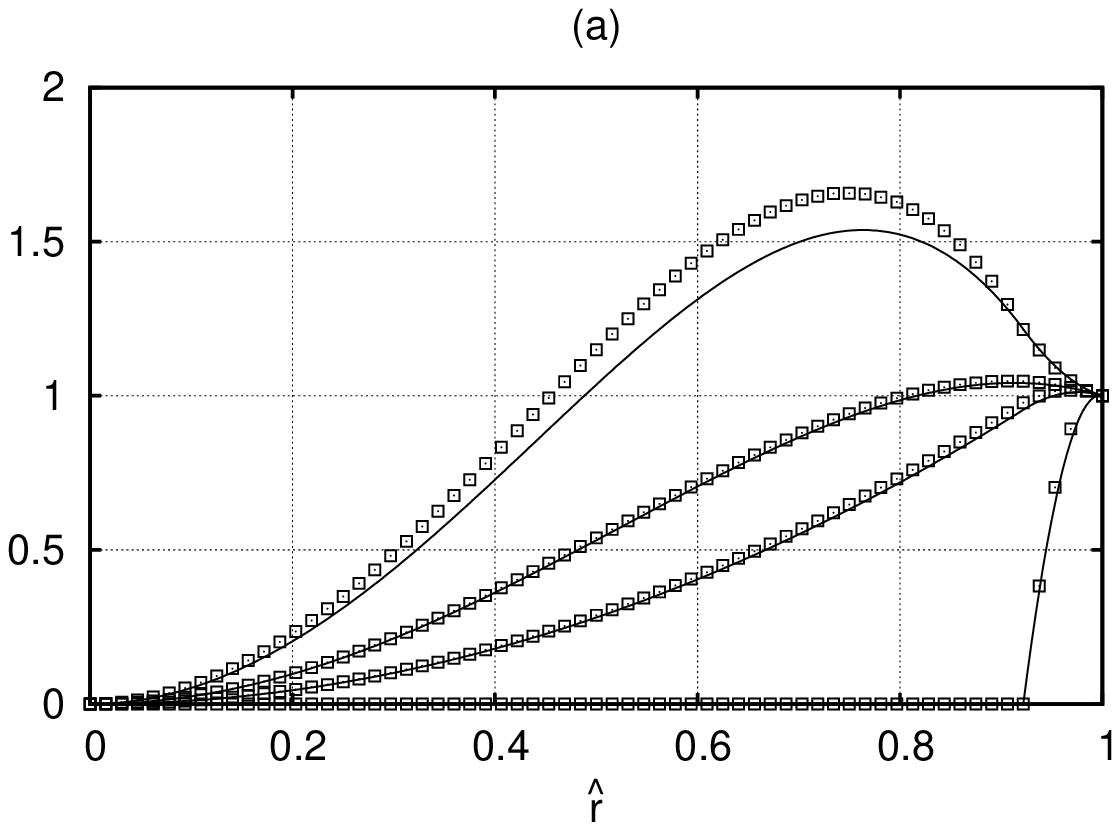}
 \includegraphics[width=8cm]{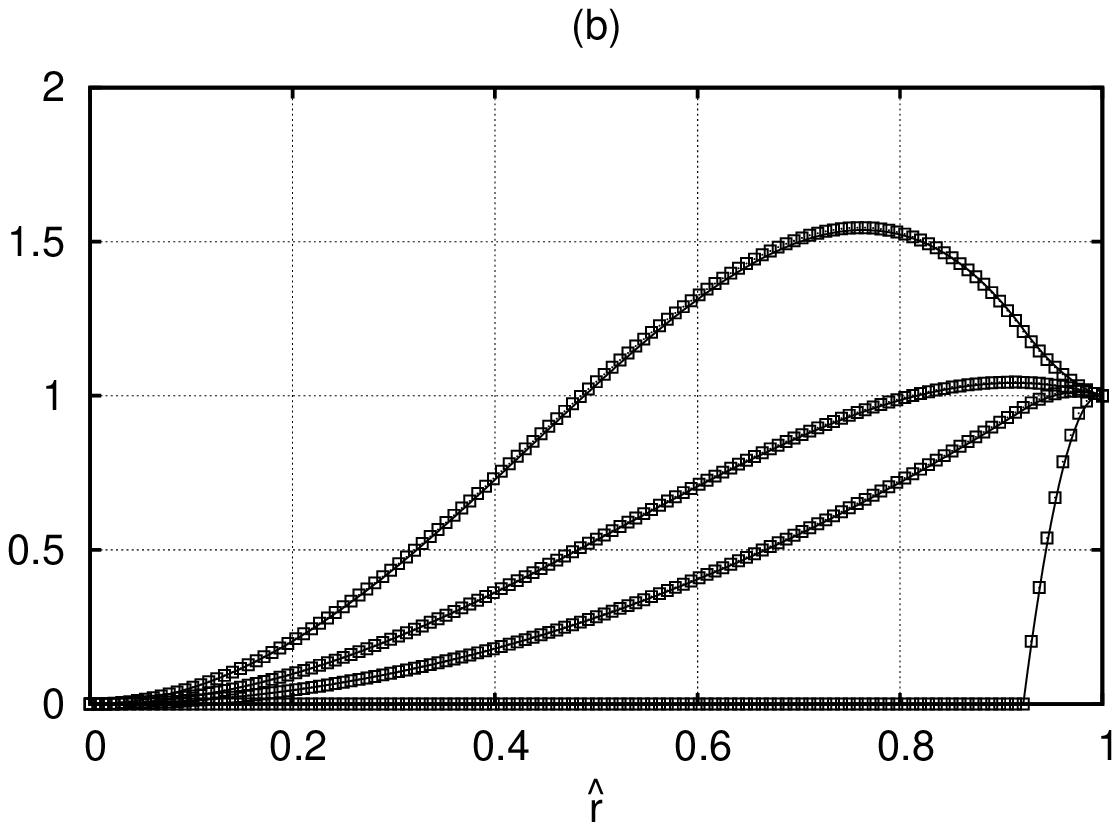}
 \caption{Numerical (points) and analytical (lines) solutions of each model. 
 Left panel(a): $N_{r1} = 65$ ($\hat{r} = [0:1]$), $N_\theta = 257$ ($\theta = [0:\pi]$) solutions. 
 Right panel(b): $N_{r1} = 129$, $N_\theta = 257$ solutions. The solutions in (b) represent
 the analytical solutions very well.}
 \label{Fig:Accuracy_check}
\end{figure*}

We checked the accuracy of the numerical solutions.
We can see the differences between 
the analytical solutions (App. \ref{App:analytical}) and numerical solutions by changing the 
mesh numbers (a:$N_{r1} = 65$, b:$N_{r1} = 129$ in Fig. \ref{Fig:Accuracy_check}),
where $N_{r1}$ refers to the mesh numbers within the stellar region ($\hat{r}$ = [0:1]).
We display the four numerical and  analytical solutions (models I, II, III and IV) in 
Fig. \ref{Fig:Accuracy_check}. As seen in Fig. \ref{Fig:Accuracy_check},
the numerical solutions in panel (b) represent the analytical solutions very well.
In actual numerical calculations which are tabulated and displayed in this paper, 
we use $N_\theta = 1025$ ($\theta = [0:\pi]$), $N_{r1} = 513$, $N_{r2} = 513$ ($\hat{r}=[1:2]$ Sec. 3.1) and
$N_{r2} = 513$, ($\hat{r}=[1:16]$ Sec. 3.2, 3.3) in order to obtain more accurate numerical results.
As for the Legendre polynomial, we sum of $P_n^1 (\cos \theta)$ to $n_{\max} = 21$ in all numerical calculations.

\section{Dissipation timescale of the current sheet at the core-crust interface}
\label{App:dissipation}

The dissipation timescale of the current sheet at the core-crust interface
is important. If  the dissipation timescale is much shorter than the Hall timescale,
our numerical result with a current sheet is not a possible configuration. 
If the value of the magnetic Reynolds number is ${\cal R}_m = 1000$ (Eq. \ref{Eq:R_m}), 
the Hall timescale is 1000 times shorter than the Ohmic dissipation timescale.
Since the Ohmic dissipation timescale is given by
\begin{eqnarray}
 t_{\rm Ohm} = \frac{4 \pi \sigma \delta r_c^2 }{c^2} \sim 5  \times 10^{7} \left(\frac{\sigma}{10^{25} \rm{s^{-1}}} \right) 
\left(\frac{\Delta r_c}{0.1 r_s}\right)^2 {\rm yr.},
\end{eqnarray}
the Hall timescale is estimated as 
\begin{eqnarray}
 t_{\rm Hall} = {\cal R}_m t_{\rm Ohm} \sim 5 \times 10^{4} {\rm yr.},
\end{eqnarray} 
where $\Delta r_c$ is a width of the crust and $r_s \sim 10^{6} {\rm cm}$
is the stellar radius.

The dissipation timescale of the current sheet depends on both
the width of the current sheet and the electrical conductivity at the interface,
but the exact value of the electrical conductivity is still not clear. 
We assumed that the value equal to that of the core electrical conductivity 
$\sigma  \sim 10^{29} \rm{s}^{-1}$ 
(\citealt{Baym_et_al_Nature_1969b}; \citealt{Neutron_Stars_1}).
If the width of the current sheet is $10^{-4} r_s$, then
the dissipation timescale of the current sheet $t_{\rm dis.}$ is evaluated by
\begin{eqnarray}
 t_{\rm dis.}
= \frac{4 \pi \sigma \Delta r_{cs}^2}{c^2} \sim 
 5 \times 10^{5} \left(\frac{\sigma }{10^{29} \rm{s}^{-1}}  \right) 
 \left(\frac{\Delta r_{cs}}{10^{-4} r_s}  \right)^2 \rm{yr.},
\end{eqnarray}
where $\Delta r_{cs}$ is the width of the current sheet.
The grid number of the star is $N_r = 513$ (see App. \ref{App:accuracy})
and the grid size is $\Delta r = r_s/512 \sim 2 \times 10^{-3} r_s$ in this paper.
Although  $\Delta r_{cs}$ is much smaller than the grid size $\Delta r$,
the dissipation timescale $t_{\rm dis.}$ is much 
larger than the Hall timescale $t_{\rm Hall}$.
Therefore,  we can treat the current sheet as a $\delta$ function (Eq. \ref{Eq:j_sur}) and
consider the magnetar models with  a current sheet {at the core-crust interface}
in this paper as possible configurations.

\end{document}